\documentclass[aps,prb,groupedaddress,amsmath]{revtex4}

\usepackage{amssymb}
\usepackage{amsmath}
\usepackage{graphicx}
\usepackage{amsbsy}
\usepackage{color}

\newcommand{\bq}{\begin{eqnarray}}
\newcommand{\eq}{\end{eqnarray}}
\newcommand{\an}[1]{#1}

\begin{document}
\title{Penetrable-Square-Well fluids: Analytical study and
Monte Carlo simulations}

\author{Riccardo Fantoni}
\email{rfantoni@unive.it}
\affiliation{Dipartimento di Chimica Fisica, Universit\`a di Venezia,
Calle Larga S. Marta DD2137, I-30123 Venezia, Italy}
\author{Achille Giacometti}
\email{achille@unive.it}
\affiliation{Dipartimento di Chimica Fisica, Universit\`a di Venezia,
Calle Larga S. Marta DD2137, I-30123 Venezia, Italy}
\date{\today}
\author{Alexandr Malijevsk\'y}
\email{amail@post.cz}
\affiliation{E. H\'ala Laboratory of Thermodynamics, Academy of
Science of the Czech Republic, Prague 6, Czech Republic and Institute
of Theoretical Physics, Department of Chemical Engineering, Imperial
College London, South Kensington Campus, London SW7 2BZ, UK }
\author{Andr\'es Santos}
\email{andres@unex.es;
URL:http://www.unex.es/fisteor/andres/}
\affiliation{Departamento de F\'isica, Universidad de Extremadura,
E-06071 Badajoz, Spain}

\begin{abstract}
We study structural and thermophysical properties of a one-dimensional
classical fluid made of penetrable spheres interacting via an attractive
square-well potential. Penetrability of the spheres is enforced by
reducing from infinite to finite the repulsive energy barrier in the
pair potentials As a consequence, an exact analytical solution is lacking even in one dimension.
Building upon previous exact analytical work in the low-density limit [Santos \textit{et al.}, Phys.\ Rev.\ E \text{77}, 051206 (2008)],
we propose an approximate theory valid at any density and in the low-penetrable regime. By comparison with specialized Monte Carlo
simulations and integral equation theories, we assess the regime of validity
of the theory. We investigate the degree of inconsistency among the various routes to thermodynamics
and explore the possibility of a fluid-fluid transition. Finally
we locate the dependence of the Fisher-Widom line on the degree
of penetrability. Our results constitute the first systematic study of penetrable
spheres with attractions as a prototype model for soft systems.
\end{abstract}

\maketitle
\section{Introduction}
\label{sec:intro}
Hard spheres constitute a paradigmatic system for many simple and
complex fluids. Steric stabilized colloids, for instance, are suspensions
made of colloidal particles coated by short linear polymers suspended
in a microscopic
solvent fluid. For sufficiently high temperature and/or in the presence of
a good solvent, those dressed colloids effectively interact as hard spheres.\cite{Barrat03}


On the other hand, a number of soft colloidal systems are always penetrable
at least to a certain extent \cite{Likos01}. 
Notable examples include for instance star-shaped \cite{Watzlawek99}
or branched-shaped \cite{Ballauff04} polymers where each macromolecue
can be roughly regarded as a sphere of a given radius (the radius of gyration),
but two particles can clearly interpenetrate to a substantially smaller distance.



A necessary (but not sufficient) condition for a one-dimensional fluid
to be a {\sl nearest-neighbor} fluid is to be a hard-core fluid, \emph{i.e.},
a fluid made of particles which cannot penetrate one another due to
the existence of an infinite repulsive potential barrier in the pair
potential $\phi(r)$.
\an{Nearest-neighbor} fluids admit an analytical exact statistical-mechanical
solution:\cite{Salsburg53} the partition function,
equation of state, and correlation functions of any order can be
calculated analytically from the knowledge of the pair potential.
This is no longer the case for non-neighbor fluids.\cite{Fantoni03}

Penetrable spheres (PS)\cite{Likos98,Malijevsky06} can be reckoned as the simplest
representation of soft colloids where the range of penetrability
can be tuned from zero (hard spheres) to infinity (ideal gas). Both limits
are amenable to an exact analytical treatment, but the intermediate case is
not.

When an attractive, short-range, square well (SW) is added to PS, one
obtains the so-called penetrable-square-well (PSW) fluid.\cite{Santos08}
On the one hand, this enriches the model so that it can also account
for short-range attractive interactions which are ubiquitous in such systems.
On the other hand, it also complicates the treatment due to
possible Ruelle instabilities associated with the lack of a well defined
thermodynamic limit.\cite{Fisher66,Ruelle69}
As the width of the well vanishes with a constant
area under the well the PSW model reduces to what we denote\cite{Santos08} as the sticky-penetrable-sphere (SPS) model. This model was
found to be thermodynamically unstable\cite{Santos08} due to the
divergence of the fourth virial coefficient. In fact, SPS model
violates the (sufficient) condition for stability (see Appendix A in Ref.\ \onlinecite{Santos08}).

We emphasize that various classes of penetrable systems have
appeared in the literature with rather different meanings.
The Widom--Rowlinson model of non-additive hard-sphere mixtures,\cite{Widom70} for instance,  is not associated with a well defined pair potential
as in the case of the present study.
Likewise, the Rikvold--Stell--Torquato ``permeable sphere'' model\cite{Torquato84,Rikvold85}
is defined through a condition on correlation function which
is not equivalent to a constant repulsive potential inside the
core region.
On the contrary, our PSW model belongs to the same class of bounded potentials as the Gaussian-core models originally
proposed by Stillnger \emph{et al.}\cite{Stillinger76}
in the late 1970s  and exploited more recently by the D\"{u}sseldorf\cite{Likos98,Lang00} and the Cambridge\cite{Louis00} groups.

In a previous paper,\cite{Santos08} we have introduced the PSW fluid model
and discussed the conditions under which the model is Ruelle stable.
In addition, we have also derived an exact low-density expansion
up to second order in the radial pair distribution function (corresponding
to the fourth order in the virial coefficient) which was shown to compete with
standard integral equation approximations such as Percus--Yevick (PY)
and hypernetted chain (HNC) over a wide region of the density-temperature
phase diagram. These exact results, however, fail to reproduce the correct
behavior when the concentration is large, due to their low-density
character.

The aim of the present paper is to extend the analysis to these more demanding conditions,
by using an approximation already successfully exploited in the PS case.
In this case it has been argued\cite{Malijevsky06} that the exact analytical solution
stemming from corresponding hard-sphere particles can be efficiently
exploited to implement a low-penetrability approximate solution (LPA, called
LTA in Ref.\ \onlinecite{Malijevsky06}). The basic idea behind the method is that
for sufficiently low penetrability, the functional form of the equations
derived in the impenetrable case can be smoothly adapted to the
penetrable case by ``healing'' a few crucial aspects of the original solution.
Building upon this idea, we here show  that this methodology can also be applied to the PSW case
by starting from the corresponding impenetrable counterpart
(\emph{i.e.}, the SW potential).

We discuss the soundness of this approximation in various ways.
First by comparing the LPA low-density results against
the exact low-density expansion which was computed in
Ref.\ \onlinecite{Santos08}. Secondly, by comparing with specialized
Monte Carlo (MC) simulations and standard integral equations
(notably PY and HNC). We show how
LPA properly describes a significant part of the phase diagram
with a performance comparable with integral equations at
a semi-analytical level.

The introduction of an attractive part in the PS potential opens the
route to some interesting questions that we also address in the
present paper. First of all, we question the existence of
a fluid-fluid phase separation in addition to the fluid-solid
transition{, by limiting our analysis within the range of
applicability of LPA, that is, we avoid densities so high that
a substantial interpenetration among particles is expected.}

Within the same LPA,  we also investigate modifications on the Fisher--Widom line,
marking the transition from oscillatory to exponential decay regimes
for correlation functions, that is known to exist even
in the SW one-dimensional fluid.\cite{Fisher69} We find an increase of the exponential decay
region and we address the physical motivations behind this.

The structure of the paper is as follows: we define the PSW model in Sec.\ \ref{sec:psw}.
In Sec.\ \ref{sec:general} we briefly recall the well known general scheme
allowing for the exact analytical solution of the class of nearest-neighbor
one-dimensional fluids. We then construct the LPA in Sec.\ \ref{sec:low}
and show how this reduces to its counterpart within the PS limit\cite{Malijevsky06}
and \an{assess} its performance in comparison with known exact results within
the low-density limit.\cite{Santos08} Sections \ref{sec:fw} and \ref{sec:eos}
contain a discussion on the Fisher--Widom line and on the routes
to thermodynamics, as predicted by the LPA, \an{respectively}. The regions in the density-temperature diagram where the LPA is only slightly thermodynamically inconsistent (and thus expected to be reliable) are discussed in Section \ref{sec:limits}, where also an improved version of the approximation is proposed. Section \ref{sec:MC_IE}
includes a very brief description on the numerical methods
(MC simulations and integral equations) discussed
in the present model. These numerical results are presented and compared with
LPA theory  in Section \ref{sec:results}. The paper ends with some concluding remarks in Sec.\ \ref{sec:conc}.
\section{The penetrable-square-well (PSW) model}
\label{sec:psw}
The PSW fluid is defined through
the following pair potential\cite{Santos08} (see Fig.\ \ref{fig:fig1}, top panel)
\begin{eqnarray} \label{sec:psw1}
\phi\left(r\right)=\left\{
\begin{array}{ll}
\epsilon_r, & r<\sigma,\\
-\epsilon_a, & \sigma<r<\sigma+\Delta,\\
0 ,          & r>\sigma+\Delta,
\end{array}\right.
\end{eqnarray}
where $\epsilon_r$ and $\epsilon_a$ are two positive constants
accounting for the repulsive and attractive parts of the potential,
respectively. The corresponding Mayer function $f(r)=e^{-\beta \phi(r)}-1$ (where $\beta=1/k_BT$ is the inverse temperature parameter) reads
\begin{eqnarray}
\label{sec:psw2}
f\left(r\right)&=& \gamma_r f_{\text{HS}}\left(r\right)+\gamma_a
\left[\Theta\left(r-\sigma\right)-
\Theta\left(r-\sigma-\Delta\right)\right]~,
\end{eqnarray}
where $\gamma_r= 1-e^{-\beta\epsilon_r}$ is the parameter measuring the
degree of penetrability varying between 0 (free penetrability) and
1 (impenetrability) and $\gamma_a= e^{\beta\epsilon_a}-1>0$ plays a similar
role for the attractive part. Here $f_{\text{HS}}(r)=\Theta(r-\sigma)-1$ is
the Mayer function for the hard-sphere model which can then
be recovered in the limit $\gamma_r \to 1$ and either $\gamma_a \to 0$ or $\Delta \to 0$.  $\Theta(r)$
is the usual step function equal to $1$ for $r>0$ and $0$ otherwise.
It also proves convenient to introduce the ratio $\gamma =
\gamma_a/\gamma_r$, which is a measure of the relative depth of the attractive well.

The above potential reduces to the corresponding PS and SW potentials in the limits $\epsilon_a \to 0$
(or $\Delta \to 0$)
and $\epsilon_r \to \infty$, respectively. Other interesting limiting
cases have already been detailed in Ref.\ \onlinecite{Santos08}.

Consider a SW fluid in one dimension: different particles
can be assigned an increasing coordinate on the  axis line
and the only possible  configurations are those
indicated with A or B
in  Fig.\ \ref{fig:fig1} (middle panel), where either the centers of two different spheres
are separated a distance greater than the attractive SW range and behave
as hard spheres (A) or they are sufficiently close
to attract each other (B). PSW spheres, on the other hand,
can interpenetrate with some energy cost so they also
display configurations such as, for instance, (C) or (D)  in Fig.\ \ref{fig:fig1}
(bottom panel). PSW fluids are then effectively a many-body problem and, as such,
not amenable to an analytical solution.

In the present paper, our analysis will be limited to the case $\epsilon_r>2\epsilon_a$
where a well defined thermodynamic limit is ensured.\cite{Santos08}

\section{General recipe for nearest-neighbor interactions }
\label{sec:general}
In this section we provide a synopsis of the main steps required by
the analytical solution of any nearest-neighbor fluid.\cite{Salsburg53,Corti98,Heying04} This
will be used in next section to introduce a motivated approximate solution
in a particular limit.
\begin{itemize}
\item {}From the Boltzman factor $e^{-\beta \phi(r)}$ compute its Laplace transform
\begin{eqnarray}
\label{sec:general1}
\widetilde{\Omega}\left(s\right) &=& \int_{0}^{\infty} dr \, e^{-s r}e^{-\beta \phi\left(r\right)}.
\end{eqnarray}
\item  The equation of state is given by
\begin{equation}
\label{sec:general2b}
\beta p =\frac{\xi}{\sigma},
\end{equation}
where $p$ is the pressure and the  parameter $\xi$  is the solution
of the equation
\begin{equation}
\label{sec:general2a}
\rho=-\frac{\widetilde{\Omega}\left({\xi}/{\sigma}\right)}{\widetilde{\Omega}^{\prime}\left({\xi}/{\sigma}\right)},
\end{equation}
where $\rho$ is the density and $\widetilde{\Omega}^{\prime}(s)=\partial \widetilde{\Omega}(s)/\partial s$.  This provides all thermodynamics.
\item The radial distribution function (RDF) can be obtained from
\begin{eqnarray}
\label{sec:general3}
\widetilde{G}\left(s\right) &=& \frac{1}{\rho} \frac{\widetilde{\Omega}\left(s+{\xi}/{\sigma}\right)}
{\widetilde{\Omega}\left({\xi}/{\sigma}\right)-\widetilde{\Omega}\left(s+{\xi}/{\sigma}\right)},
\end{eqnarray}
which is the Laplace transform of the RDF $g(r)$.
\end{itemize}
This is sufficient to compute both thermodynamics and structural properties of any
one-dimensional system with nearest-neighbor interactions.

At odds to this class of problems, penetrable spheres do not possess any
analytical solution even in one dimension. This is because it is not possible
to convolute appropriate Laplace transform along a one-dimensional axis, which
is the essential feature  rendering the short-range one-dimensional models
solvable. In turn this is due to the existence of multiple ``blobs'' formed
by interpenetrating spheres so that it is no longer possible to
``order'' them along a line in such a way that they do not cross each other,
a key point to the existence of the analytical solution (see Fig.\ \ref{fig:fig1},
middle panel).

Because of this, we now turn our attention to a  motivated approximation
which amounts to assume a slight decrease from an infinite repulsive
barrier, an approximation which will be denoted as low-penetrability.
\section{The low-penetrability approximation (LPA)}
\label{sec:low}

\subsection{Construction of the approximation}
In Ref.\ \onlinecite{Santos08} we have followed the philosophy of considering
a low-density expansion to provide exact analytical results valid
up to second order in the RDF  $g(r)$ and up to
fourth order in the virial expansion. This is a very useful
exact limit case to test  approximate theories and
numerical simulations, but it has the considerable disadvantage \an{of being} limited to very low densities. We now consider a different approach
where density can in principle be arbitrarily large but we assume low penetrability
among different spheres, patterned after a similar
idea already used in the PS case.\cite{Malijevsky06}

For notational simplicity, in the following, lengths will be measured in units of $\sigma$
(so that $\sigma=1$) and we introduce  $\lambda=1+\Delta/\sigma$ as a dimensionless measure
of the external well boundary.

The Laplace transform of the Boltzmann factor $e^{\beta \phi (r)}$ for the PSW model is
\begin{eqnarray}
\label{sec:low1}
\widetilde{\Omega}\left(s\right)&=&\frac{1-\gamma_r}{s}+
\frac{\gamma_r}{s} \left[\left(1+\gamma\right)e^{-s}- \gamma e^{-\lambda s}\right].
\end{eqnarray}
The PSW fluid is not a nearest-neighbor fluid, as remarked, but it reduces to the nearest-neighbor SW fluid
as $\gamma_r\to 1$ (and $\Delta<1$). In this limit, it is natural to
use the recipe given in Sec.\ \ref{sec:general} for  the SW fluid, to derive an approximate
equation of state  and an approximate $g(r)$ from Eqs.\ (\ref{sec:general2b}), (\ref{sec:general2a}), and
(\ref{sec:general3}), respectively. This, however, must be exercised with care as important general properties
of  any model, such as for instance the continuity of the cavity function $y(r)=g(r) e^{\beta \phi(r)}$,
are typically lost by this brute force procedure. The driving idea behind this simple
low-penetrability approximation (LPA) is then to keep the general features
of the original SW solution and enforce some specific modifications guided by
the accounting of increasingly important constraints.\cite{Note2}

Our LPA implementation amounts to replacing Eq.\ (\ref{sec:general3}) with
\begin{eqnarray}
\label{sec:low2}
\widetilde{G}\left(s\right) &=& \frac{1}{\rho} \frac{\widetilde{\Omega}\left(s+\zeta \right)}
{\widetilde{\Omega}_{0}\left(\zeta \right)-\widetilde{\Omega}_{0}\left(s+\zeta \right)},
\end{eqnarray}
where $\widetilde{\Omega}_{0}(s)$ is (formally) the Laplace transform of the Bolzmann factor of the
SW model which can be obtained from  $\widetilde{\Omega}(s)$ by discarding
 the first term on the right-hand-side of Eq.\ (\ref{sec:low1}), i.e.,
 \begin{eqnarray}
\label{sec:low1bis}
\widetilde{\Omega}_0\left(s\right)&=&
\frac{\gamma_r}{s} \left[\left(1+\gamma\right)e^{-s}- \gamma e^{-\lambda s}\right].
\end{eqnarray}
This simple choice can be shown to be fully equivalent to keeping Eq.\ (\ref{sec:general3})
but with a replacement $e^{-s} \to e^{-(s-a)}$ in Eq.\ (\ref{sec:low1}), where the free parameter
$a$ is fixed by the continuity condition of the cavity function $y(r)$
at the hard-core discontinuity $r=1$. This is known to be the most important
feature to obtain a correct representation in integral equation theories of SW fluids,
both from the analytical and the numerical viewpoint.\cite{Yuste93,Santos06,Giacometti09}

We note that, unlike the SW counterpart, $\zeta \ne \beta p$. It is a transcendental function
of $\beta$ and $\rho$ which can be obtained by ensuring the correct behavior
of $g(r)\to 1$ as $r \to \infty$ or, equivalently, $s\widetilde{G}(s) \to 1$ as
$s \to 0$.
{}From Eq.\ (\ref{sec:low2}), this gives
\begin{eqnarray}
\label{sec:low3}
\rho &=& - \frac{\widetilde{\Omega}\left(\zeta\right)}{\widetilde{\Omega}_0^{\prime} \left(\zeta\right)}=
 \zeta\frac{h+1-q}{1-q+(1-\lambda q)\zeta},
\end{eqnarray}
where in the second equality we have introduced the following quantities:
\begin{eqnarray}
\label{sec:low4}
q&=& \frac{\gamma}{1+\gamma} e^{-\zeta \Delta} ,\\
h&=& \frac{1-\gamma_r}{\gamma_r\left(1+\gamma\right)} e^{\zeta}.
\label{sec:low5}
\end{eqnarray}
For  given values of the potential parameters ($\Delta$, $\epsilon_r$, and $\epsilon_a$) and for given values of the inverse temperature
$\beta$ and the auxiliary parameter $\zeta$, the quantities $q$ and $h$ are obtained from Eqs.\ (\ref{sec:low4}) and (\ref{sec:low5})
and inserted into Eq.\ (\ref{sec:low3}) to determine the density $\rho$. The impenetrable SW potential corresponds to
the limit $h \to 0$.

In order to compute the RDF $g(r)$ we first compute explicitly $\widetilde{G}(s)$ from Eqs.\ (\ref{sec:general3}) and (\ref{sec:low1})
\begin{eqnarray}
\label{sec:low6}
\widetilde{G}\left(s\right)&=&\frac{1}{\rho}
\frac{h+e^{-s}\left(1-q e^{-s\Delta}\right)}{\left(1-q\right)\left(1+{s}/{\zeta}\right)-e^{-s}\left(1-q e^{-s\Delta}\right)}.
\end{eqnarray}
Upon expanding the denominator in Eq.\ (\ref{sec:low6}) in powers of $(1-q e^{-s \Delta})(1+s/\zeta)$, and inverting the Laplace
transform term by term one gets
\begin{eqnarray}
\label{sec:low7}
\rho g\left(r\right)=\frac{h\zeta}{1-q} e^{-\zeta r}+\sum_{n=1}^\infty\sum_{k=0}^n \binom{n}{k}(-q)^k \psi_{n}\left(r-n-k\Delta\right)
\Theta\left(r-n-k\Delta\right),
\end{eqnarray}
where
\begin{eqnarray}
\label{sec:low8}
\psi_{n}(r)=\left(\frac{\zeta}{1-q}\right)^n\left[\frac{r^{n-1}}{(n-1)!}+\frac{h\zeta}{1-q}\frac{r^n}{n!}\right] e^{-\zeta r}.
\end{eqnarray}

We anticipate that the LPA does not capture correctly the  $r<\Delta$ trend at high densities, while it works well for
$r>\Delta$. The reason for this can be traced back to the failure of the LPA to account for the discontinuous slope of
 the cavity function $y(r)$ at $r=\Delta$. Moreover, \an{the appproximate} $y(r)$ turns out to
be discontinuous rather than continuous at $r=\lambda$,
as detailed in Appendix \ref{app:appa}. These deficiencies can be accounted for
step by step at the price of an increase in the complexity of the approximation,
and are a consequence of the phenomenological nature of the LPA.
This will be further discussed in Sec.\ \ref{sec:limits}.

As already remarked, the PSW model reduces in the appropriate limit to the
penetrable analogue of Baxter's sticky hard spheres, denoted as
SPS in Ref.\ \onlinecite{Santos08}. This is further
elaborated in Appendix \ref{app:appb}, where it is also discussed the
LPA  of the SPS model. We have explicitly checked this
is indeed the limit for PSW in the limit of very narrow and very deep
well. On the other hand, we have also found (see Appendix \ref{app:appb}) that this model
is also thermodynamically unstable as it violates the
stability criterion $\epsilon_r > 2 \epsilon_a$, and hence it will \an{not be} further discussed in the
remaining of this paper.
\subsection{The penetrable-rod limit}
\label{subsec:penetrable}

Here we show that either  in the limit
$\epsilon_a\to 0$ (which implies $\gamma\to 0$) or, alternatively, in the limit $\Delta\to 0$, the LPA that
we just found for the PSW model reduces to the corresponding one
proposed in Ref.\ \onlinecite{Malijevsky06} for the PS model.

Taking the limit $\gamma\to 0$  in Eq.\ (\ref{sec:low1}) one finds
Eq.\ (2.53) of Ref.\ \onlinecite{Malijevsky06}. Moreover $q\to 0$ and
$h\to (\gamma_r^{-1}-1)e^\zeta$ and so Eq.\ (\ref{sec:low3}) reduces
to $\rho=[1+(\gamma_r^{-1}-1)e^\zeta]/(1+\zeta^{-1})$ which can be rewritten as
$(\xi-\zeta)e^{-\zeta}/(\gamma_r^{-1}-1)=\zeta$ with $\xi=\rho(1+\zeta)$, which coincides with
Eq.\ (4.4) of Ref.\ \onlinecite{Malijevsky06} where our $\zeta$ replaces their
$\xi^\prime$.

It is straightforward to check that the same expressions for $\widetilde{\Omega}(s)$ and for $\rho$ in terms of $\zeta$ and $\gamma_r$
are obtained in the alternative limit $\Delta\to 0$. Hence LPA for PS is fully recovered.

\subsection{Comparison with exact low-density expansion}
\label{subsec:comp}
It proves interesting to compare the LPA to order $\rho$ with the exact results derived
in Ref.\ \onlinecite{Santos08} based on a low-density expansion, in order to
assess the ability of LPA to reproduce low-density results.

The general expansion of $g(r)$ in powers of the density $\rho$ has the following structure\cite{Hansen86}
\begin{eqnarray}
\label{subsec:comp1}
g\left(r\right)=g_0\left(r\right)+g_1\left(r\right)\rho+\cdots.
\end{eqnarray}
The exact results for $g_0(r)$ and $g_1(r)$ have been derived in Ref.\ \onlinecite{Santos08}:
\begin{eqnarray}
\label{subsec:comp2}
g_0^{\text{exact}}\left(r\right)= \begin{cases} 1-\gamma_r,&r<1,\\
1+\gamma\gamma_r,&1<r<1+\Delta,\\
1,&r>1+\Delta,
\end{cases}
\end{eqnarray}
\begin{eqnarray}
\label{subsec:comp3}
g_1^{\text{exact}}\left(r\right)=\gamma_r^2\begin{cases} (1-\gamma_r)\left[2\left(1+\gamma^2\Delta\right)-
r\left(1+2\gamma+2\gamma^2\right)\right],&0\leq r\leq\Delta,\\
(1-\gamma_r)\left(2-2 \gamma\Delta-r\right),  & \Delta \leq r< 1, \\
(1+\gamma\gamma_r)\left(2-2 \gamma\Delta-r\right),  & 1< r< 1+\Delta, \\
2-2 \gamma\Delta-r,  & 1+\Delta< r\leq 2, \\
\gamma\left(2+\gamma\right)\left(r-2\right)-2\gamma\Delta, & 2\leq r\leq 2+\Delta,  \\
\left(2+2\Delta -r\right)\gamma^2, & 2+\Delta \leq r\leq 2+2\Delta,  \\
 0, & 2+2\Delta \leq r.
\end{cases}
\end{eqnarray}

In order to compare $g_0^{\text{exact}}(r)$ and $g_1^{\text{exact}}(r)$ with LPA results, we
expand $\zeta$ as derived from Eq.\ (\ref{sec:low3}) to lowest order in density,
$\zeta= \zeta_0 \rho+\zeta_1 \rho^2 + \mathcal{O}(\rho^3)$,
and plug the results into Eqs.\ (\ref{sec:low7}) and (\ref{sec:low8}).
This yields Eq.\ (\ref{subsec:comp1}), where the coefficients $g_0(r)$ and $g_1(r)$
are computed within the LPA. Whereas
$g_0(r)=g_0^{\text{exact}}(r)$, $g_1(r)$ is found to differ from the exact result
$g_1^{\text{exact}}(r)$. Analytical expressions
for $\zeta_0$, $\zeta_1$, and $g_1(r)$ can be found in Appendix \ref{app:appc}.

Having done this, one can estimate the difference in the cavity function
between LPA and exact results to order $\rho$, which reads (see Appendix \ref{app:appc})
\begin{eqnarray}
\label{subsec:comp4}
y_1^{\text{exact}}(r)-y_1(r)=\begin{cases}
C_1+D_1\Delta+E_1(\Delta-1)+F_1(r-\Delta),&0\leq r\leq \Delta,\\
C_1+D_1\Delta+E_1(r-1),&\Delta\leq r\leq 1,\\
C_1+D_1(1+\Delta-r),&1\leq r\leq 1+\Delta,
\end{cases}
\end{eqnarray}
where
\begin{eqnarray}
\label{subsec:comp5a}
C_1&=&\gamma\gamma_r^2(1-\gamma_r)\frac{1+\gamma}{1+\gamma\gamma_r}\Delta,\\
\label{subsec:comp5b}
D_1&=&\gamma\gamma_r\frac{(1-\gamma_r)^2}{1+\gamma\gamma_r},\\
\label{subsec:comp5c}
E_1&=&\gamma_r(1-\gamma_r), \\
\label{subsec:comp5d}
F_1&=&\gamma_r[1-\gamma_r-2\gamma\gamma_r(1+\gamma)].
\end{eqnarray}

The right-hand side of Eq.\ (\ref{subsec:comp4}) preserves the continuity of $y_1(r)$ at $r=\Delta$ and $r=1$,
 but imposes
the continuity of $y_1(r)$ at $r=1+\Delta$ and that of $y_1'(r)$ at $r=1$ and $r=1+\Delta$, as well as the discontinuity of the exact $y_1'(r)$ at $r=\Delta$.
The latter discontinuity is, according to Eqs.\ \eqref{subsec:comp2} and \eqref{subsec:comp3},
\begin{eqnarray}
\label{subsec:comp6}
\lim_{\rho\to 0}\frac{y'(\Delta^+)-y'(\Delta^-)}{\rho y(\Delta)}=2\gamma\gamma_r^2({1+\gamma}).
\end{eqnarray}

\section{The Fisher-Widom line}
\label{sec:fw}
In a remarkable piece of work,\cite{Fisher69} Fisher and Widom
argued that the asymptotic decay
of the correlation functions is determined by the nature of the poles
$s_i=s_i(\beta,\rho)$ $(i=1,2,3,\ldots)$, with largest real
part, of the Laplace transform $\widetilde{G}(s)$ of the RDF. This asymptotic decay can be of two different types: oscillatory at
high densities and/or high  pressures, and monotonic for low
densities and/or pressures. The latter regime can exist only in the
presence of competing effects in the potential function, so it cannot exist
for purely repulsive short-range potentials, such as HS and PS potentials.

In particular, rather general arguments\cite{Evans93} suggest a behavior
\begin{eqnarray}
\label{sec:fw1}
g\left(r\right)-1&=& \sum_{i} A_{i} e^{s_i r}
\approx A_1 e^{s_1 r},
\end{eqnarray}
where we have specialized to one-dimensional systems and the sum runs over the discrete sets of poles $s_i$, $A_i$
being (in general complex) amplitudes. The asymptotic behavior
of $g(r)$ is dominated by the pole $s_1$ having the least negative
real part (to ensure stability of the liquid). If $s_1$ is complex, its conjugate $s_2=s_1^*$ must also be included in the asymptotic behavior.

Fisher and Widom derived the line ---henceforth denoted as
Fisher--Widom (FW) line--- where
this transition takes place, both in the pressure versus temperature
and in the density versus temperature diagrams, for the one-dimensional SW potential.
On crossing this line, one finds a sharp transition in the
character of the RDF $g(r)-1$: For any fixed temperature
in the $p$-$T$ plane, $g(r)$ has an oscillatory character above the FW line
and an exponential decay below it. The transition is a signature
of local ordering without any singularities in thermodynamical
quantities as there is no phase transition in the one-dimensional SW fluid.
In three-dimensions, the FW line precedes the coexistence line when
lowering the pressure at a fixed temperature. This has been numerically
observed for various fluids including SW,\cite{Evans93} Lennard--Jones,\cite{DeCarvalho94,Vega95,Dijkstra00}
 and other softer potentials.\cite{Tarazona03}

In view of the possibility for PSW to display fluid-fluid
and  fluid-solid phase transitions in spite of their one-dimensional
character, it is interesting to wonder what happens to the FW
line in the transition from SW to PSW. We now analyze this in
the framework of the LPA.

The poles of $\widetilde{G}(s)$
(different from $s=0$) can be read off from Eq.\ (\ref{sec:low2}):
\begin{eqnarray}
\label{sec:fw2}
\widetilde{\Omega}_0\left(s+\zeta\right)=\widetilde{\Omega}_0\left(\zeta\right)~.
\end{eqnarray}
As we are here interested in the pole with the negative real part closest to the
origin we set $s=-x\neq 0$ as the real root of Eq.\ (\ref{sec:fw2}),
\begin{eqnarray}
\label{sec:fw3}
\widetilde{\Omega}_0\left(\zeta-x\right)=\widetilde{\Omega}_0\left(\zeta\right)~,
\end{eqnarray}
and $s=-x^\prime\pm iy$ as the complex root with the least negative
real part, i.e.,
\begin{eqnarray}
\label{sec:fw4}
\text{Re}\, \widetilde{\Omega}_0\left(\zeta-x^\prime\pm iy\right)&=&\widetilde{\Omega}_0\left(\zeta\right)~,\\
\label{sec:fw5}
\text{Im} \,\widetilde{\Omega}_0\left(\zeta-x^\prime\pm iy\right)&=&0~.
\end{eqnarray}
The pole $s_1$ determining the asymptotic behavior is either $s_1=-x$ (monotonic decay) if $x<x'$ or $s_1=-x'\pm y$ (oscillatory decay) if $x>x'$.
The FW transition takes place when $x=x'$.

Equation (\ref{sec:fw3}) yields the condition
\begin{eqnarray}
\label{sec:fw6}
e^{-s}(1-qe^{-s\Delta})&=&(1-q)\left(1+\frac{s}{\zeta}\right)~,
\end{eqnarray}
where  $q$ is given by Eq. (\ref{sec:low4}).
Quite interestingly, as the parameter $h$ is missing, this equation formally coincides
with its SW counterpart, originally studied by Fisher and Widom [see Eq.\ (3.6) in Ref.\ \onlinecite{Fisher69}].
We can rewrite Eqs.\ (\ref{sec:fw3})--(\ref{sec:fw5}) as follows:
\begin{eqnarray}
\label{sec:fw7}
e^x\left(1-qe^{x\Delta}\right)=(1-q)\left(1-\frac{x}{\zeta}\right),\\
\label{sec:fw8}
e^{x'}\left(\cos y-qe^{x'\Delta}\cos\lambda y\right)=(1-q)\left(1-\frac{x'}{\zeta}\right),\\
\label{sec:fw9}
e^x\left(\sin y-qe^{x\Delta}\sin\lambda y\right)=-(1-q)\frac{y}{\zeta}.
\end{eqnarray}
At the FW transition ($x=x'$), Eqs.\ (\ref{sec:fw7})--(\ref{sec:fw9})
 form a set of three coupled equations whose solution yields $x$, $y$, and $\zeta$ as functions of $q$.
Use of Eq.\ (\ref{sec:low3}) then gives the line in the $\rho$-$T$ plane.

It proves convenient to eliminate $\zeta$ from Eqs.\ (\ref{sec:fw7}) and (\ref{sec:fw8})
to obtain
\begin{eqnarray}
\label{sec:fw10}
x=\frac{1}{\Delta}\ln\frac{1-\cos y}{q(1-\cos\lambda y)},
\end{eqnarray}
so that from Eqs.\ (\ref{sec:fw7}) and (\ref{sec:fw9}) we can now get
\begin{eqnarray}
\label{sec:fw11}
\zeta=x-y\frac{\cos y-\cos\lambda y}{\sin y-\sin\lambda y+\sin y \Delta }.
\end{eqnarray}
When Eq.\ (\ref{sec:fw10}) and (\ref{sec:fw11}) are inserted into Eq.\ (\ref{sec:fw7}) we get
\begin{eqnarray}
\label{sec:fw12}
\sin y-\sin\lambda y+\sin y\Delta-\frac{y}{x}\left(\cos y-\cos \lambda y\right)=-e^{-x}(1-q)\frac{y}{x}\left(1-\cos\lambda y\right),
\end{eqnarray}
\noindent
where $x(q,y)$ is given by Eq.\ (\ref{sec:fw10}) so that this is a transcendental equation in $y(q)$.
Once $y(q)$ is known from Eq.\ (\ref{sec:fw12}), Eqs.\ (\ref{sec:fw10}) and (\ref{sec:fw11}) provide $x(q)$ and $\zeta(q)$, respectively.
The parameter $\gamma(q)$ is obtained by inverting  Eq.\ (\ref{sec:low4}),
\begin{eqnarray}
\label{sec:fw13}
\gamma(q)=\frac{q}{e^{-\zeta(q)\Delta}-q},
\end{eqnarray}
and the inverse temperature $\beta(q)$ is obtained from
\begin{eqnarray}
\label{sec:fw14}
\gamma(q)=\frac{e^{\beta(q)\epsilon_a}-1}{1-e^{-\beta(q)\epsilon_r}}
\end{eqnarray}
on using the definitions of $\gamma$, $\gamma_r$, and $\gamma_a$.

Finally, Eqs.\ (\ref{sec:low3}) and (\ref{sec:low5}) provide $\rho(q)$ and the combination of $\beta(q)$ and $\rho(q)$ yields the  FW line in the
$\rho$-$T$ plane. In order to have it in the $p$-$T$ plane one needs to get before the equation of state
and the result will depend on the chosen route (virial, compressibility, or energy). This is discussed  in the following section.
\section{Equation of state }
\label{sec:eos}
As PSW is not an exactly solvable model, thermodynamics will in general depend
upon the followed route, so we are going to check the three standard routes
(virial, compressibility, and energy) for the compressibility factor
$Z=\beta p/\rho$, as predicted by the LPA.

The virial route is defined by
\begin{eqnarray}
Z&=&1-\rho \beta\int_{0}^{\infty} dr \, r y\left(r\right) e^{-\beta \phi\left(r\right)}  \phi^{\prime}\left(r\right)
\label{sec:eos1}
\end{eqnarray}
which, using standard manipulations,\cite{Hansen86} yields
\begin{eqnarray}
\label{sec:eqs2}
Z&=&1+\rho\gamma_r[(1+\gamma)y(1)-\gamma\lambda y(\lambda)].
\end{eqnarray}
As $y(\lambda^-)\neq y(\lambda^+)$ within LPA (see Appendix \ref{app:appa}),
$y(\lambda)= \frac{1}{2}\left[y(\lambda^-)+y(\lambda^+)\right]$ is assumed.
Thus, using Eqs.\ (\ref{app:a4}) and (\ref{app:a8}) we get
\begin{eqnarray}
\label{sec:eos3}
Z&=&1+\frac{\zeta}{1-q}\left\{1-\lambda q\left[1+{\gamma_r(1+\gamma)}\frac{1+\gamma\gamma_r/2}{1+\gamma\gamma_r}\frac{h\zeta\Delta}{1-q}\right]\right\}.
\end{eqnarray}
It is easy to check using Eqs.\ (\ref{sec:low3}) and (\ref{sec:eos3}) that in the case of the SW model ($h=0$) one recovers the
expected result $Z=\zeta/\rho$.

Next we consider the compressibility route:
\begin{eqnarray}
\label{sec:eos4}
\chi\equiv \frac{1}{\beta}\left(\frac{\partial\rho}{\partial p}\right)_\beta&=&1+2\rho\int_0^\infty dr\, [g\left(r\right)-1] \nonumber \\
&=&1+2\rho\lim_{s\to 0}\left[\widetilde{G}\left(s\right)-s^{-1}\right].
\end{eqnarray}
Using Eqs.\ (\ref{sec:low3}) and (\ref{sec:low6}) the \an{last  term} of Eq.\ (\ref{sec:eos4})
can be written as
\begin{eqnarray}
\label{sec:eos5}
2\rho\lim_{s\to 0}\left[\widetilde{G}\left(s\right)-s^{-1}\right]
&=&\left(\frac{\partial\rho}{\partial\zeta}\right)_\beta-\frac{{\widetilde{\Omega}}'(\zeta)}{{\widetilde{\Omega}}_0'(\zeta)}.
\end{eqnarray}
Introducing the quantity
\begin{eqnarray}
\label{sec:eos6}
X(\zeta)\equiv \frac{1}{(\partial\rho/\partial\zeta)_\beta}
\frac{{\widetilde{\Omega}}_0'(\zeta)-{\widetilde{\Omega}}'(\zeta)}{{\widetilde{\Omega}}_0'(\zeta)},
\end{eqnarray}
Eq.\ \eqref{sec:eos4} becomes
\begin{eqnarray}
\label{sec:eos7}
\chi=\left(\frac{\partial\rho}{\partial\zeta}\right)_\beta\left[1+X(\zeta)\right] ,
\end{eqnarray}
and using the definition of $\chi$ we find
\begin{eqnarray}
\label{sec:eos8}
\beta\left(\frac{\partial p}{\partial\zeta}\right)_\beta&=&\frac{1}{1+X(\zeta)}.
\end{eqnarray}
Therefore the compressibility route yields
\begin{eqnarray}
\label{sec:eos9}
\beta p(\zeta)&=&\int_0^\zeta\frac{d\zeta'}{1+X(\zeta')}.
\end{eqnarray}
In the SW limit one clearly has $X(\zeta)=0$ and $\beta p=\zeta$, as it should.

The energy route is by far the most delicate. We start from the internal energy per particle
\begin{eqnarray}
\label{sec:eos10}
u&=&\frac{1}{2\beta}+\rho\int_0^\infty dr\, \phi(r) g(r) \nonumber \\
&=&\frac{1}{2\beta}+\epsilon_r\rho\int_0^1 dr\,  g(r)-\epsilon_a\rho\int_1^{\lambda} dr\,  g(r).
\end{eqnarray}
Equation (\ref{app:a1})  provides the necessary result for $g(r)$ in the interval $0<r<\lambda$, so that
\begin{eqnarray}
\label{sec:eos11}
u&=&\frac{1}{2\beta}+\epsilon_r \frac{h}{1-q}\left(1-e^{-\zeta}\right)
  -\epsilon_a \left[\frac{1}{1-q}\left(1-e^{-\zeta\Delta}\right)\left(1+\frac{h}{1-q}+h e^{-\zeta}\right)
  - \frac{h}{(1-q)^2}e^{-\zeta\Delta}\zeta\Delta \right].
\end{eqnarray}
In order to obtain $\beta p$ from $u$ we exploit the following thermodynamic relation
\begin{eqnarray}
\label{sec:eos12}
\rho^2\left(\frac{\partial u}{\partial\rho}\right)_\beta&=&\left(\frac{\partial \beta p}{\partial\beta}\right)_\rho
\end{eqnarray}
and the  identity
\begin{eqnarray}
\label{sec:eos13}
\left(\frac{\partial u}{\partial\rho}\right)_\beta&=&
\left(\frac{\partial u}{\partial\zeta}\right)_\beta\left(\frac{\partial \zeta}{\partial\rho}\right)_\beta
\end{eqnarray}
to obtain
\begin{eqnarray}
\label{sec:eos14}
\left(\frac{\partial \beta p}{\partial\beta}\right)_\rho&=&
\frac{\rho^2}{\left({\partial\rho}/{\partial \zeta}\right)_\beta}\left(\frac{\partial u}{\partial\zeta}\right)_\beta.
\end{eqnarray}
Once again one can check that Eq.\ (\ref{sec:eos14}) is satisfied by the SW result  $\beta p=\zeta$.

The right-hand side of Eq.\ (\ref{sec:eos14}) is a function of $\beta$ and $\rho$, which we denote as $R(\beta,\rho)$,
as $\zeta$ is itself a function of the same variables through Eq.\ (\ref{sec:low3}).
Thus, Eq.\ (\ref{sec:eos14}) gives
\begin{eqnarray}
\label{sec:eos15}
\beta p(\beta,\rho)&=&\zeta(\rho,\beta_\text{max})-\int_\beta^{\beta_\text{max}} d\beta'\, R(\beta',\rho),
\end{eqnarray}
where  $\beta_\text{max}$ is a conveniently chosen high value.\cite{Note3}

\section{Reliability of LPA and possible improvements}
\label{sec:limits}
We are now in the position to draw a qualitative phase diagram
 in  the $\rho\sigma$-$k_B T/\epsilon_a$ plane
indicating the boundary where the LPA can be approximately regarded to
be reliable. Of course, a definite reliability test is only possible after comparison with computer simulation results but before that we can use the internal  consistency among the three thermodynamic routes as a reliability criterion.

In general, it turns out that thermodynamic inconsistency increases as the temperature and the density increase.
To characterize this, let us define a density $\rho_\text{lim}(T)$ such that the largest relative deviation among the three routes  is smaller than 5\% if $\rho<\rho_\text{lim}(T)$. Therefore, all the points in the temperature-density plane with $\rho<\rho_\text{lim}(T)$ represent states where the LPA is only weakly inconsistent. This boundary line is shown in Fig.\ \ref{fig:fig2} for three representative cases {of the pair
($\epsilon_r/\epsilon_a$  and $\Delta$)}. We observe that the region where the LPA is thermodynamically consistent shrinks as $\epsilon_r/\epsilon_a$ decreases and/or $\Delta$ increases. In any case, it is noteworthy that if the density is smaller than a certain value (which of course depends on $\epsilon_r/\epsilon_a$  and $\Delta$), the LPA remains thermodynamically consistent even for high temperatures.

The above reliability criterion is based on thermodynamics and thus it is a global one. On the other hand, we know that the LPA has some local shortcomings, such as an artificial discontinuity of the cavity function at the  point $r=1+\Delta$, as shown in Appendix \ref{app:appa}. Moreover, it does not predict the discontinuity of the slope of the RDF at $r=\Delta$, already present by the exact result to first order in density, as indicated by Eq.\ \eqref{subsec:comp6}.

As anticipated in Sec.\ \ref{sec:low}, we can extend the validity
of the LPA  by a suitable modification of the
cavity function $y(r)$ in order to ensure a correct behavior
both within the core region and at the well-edge discontinuity.
We outline a possible approach to this issue in the remainder of this Section.

{Inspired by the comparison with exact low-density results as given in Section \ref{subsec:comp},
we } modify the LPA  (mLPA) by adding linear terms in the region $0\leq r\leq 1+\Delta$,
following a form based on that of Eq.\ \eqref{subsec:comp4}, namely
\begin{eqnarray}
\label{sec:limit1}
 g^{\text{mLPA}}(r)&=& g(r)+\frac{1}{\rho}
\begin{cases}
(1-\gamma_r)[C+D\Delta+E(\Delta-1)+F(r-\Delta)],&0\leq r\leq \Delta,\\
(1-\gamma_r)[C+D\Delta+E(r-1)],&\Delta\leq r< 1,\\
(1+\gamma\gamma_r)\left[C+D(1+\Delta-r)\right],&1< r< 1+\Delta,
\end{cases}
\end{eqnarray}
where $g(r)$ is the LPA radial distribution function as given Eq.\ (\ref{sec:low7}). The
 parameters $C$, $D$, and $E$ can be determined by imposing the continuity of $y(r)$ at $r=1+\Delta$ and of $y'(r)$ at $r=1+\Delta$ and $r=1$, respectively.
They are given by
\begin{eqnarray}
\label{sec:limit2}
C&=&\frac{\gamma_r(1+\gamma)}{1+\gamma\gamma_r}\frac{hq\zeta^2}{(1-q)^2}\Delta, \\
\label{sec:limit3}
D&=&\frac{1-\gamma_r}{1+\gamma\gamma_r}\frac{hq\zeta^2}{(1-q)^2}+{C}{\zeta},\\
\label{sec:limit4}
E&=&\frac{1}{1+\gamma\gamma_r}\frac{h\zeta^2}{(1-q)^2}-D.
\end{eqnarray}
The addition of the coefficient $F$ is motivated by  the exact results
to first order in density, Eq.\ \eqref{subsec:comp3}, showing that, as recalled above, $g(r)$ exhibits a change of slope at $r=\Delta$,
a feature not accounted
for by the LPA. In order to determine the coefficient $F$ we extend the exact low-density condition \eqref{subsec:comp6}
to finite density. This implies
\begin{equation}
  F=E-2\rho\gamma\gamma_r^2(1+\gamma)\left[\frac{1+\gamma}{\gamma(1-\gamma_r)}\frac{hq\zeta }{1-q}+C+D\Delta+E(\Delta-1)\right].
  \label{sec:limit5}
\end{equation}
It is straightforward to check that $C=C_1\rho^2+\mathcal{O}(\rho^3)$, $D=D_1\rho^2+\mathcal{O}(\rho^3)$,
$E=E_1\rho^2+\mathcal{O}(\rho^3)$, $F=F_1\rho^2+\mathcal{O}(\rho^3)$, where $C_1$, $D_1$, $E_1$, and $F_1$
are given by Eqs.\ (\ref{subsec:comp5a})--(\ref{subsec:comp5d}).
Therefore, the  mLPA is exact to first order in density.

The discussed modification of LPA then takes care of the continuity of the cavity function
$y(r)$ at both interaction discontinuities $r=1$ (already accounted for within LPA) and $r=\lambda$
(where the original LPA fails to provide continuity), and it correctly matches the
exact results for $g(r)$ up to first order in density. A similar modification
of the SPS model, discussed in Appendix \ref{app:appa}, would heal the
discontinuity appearing in the corresponding LPA values $y^{\text{SPS}}(1^{+}) \ne
y^{\text{SPS}}(1^{-})$, which is a consequence of the combined effects
 of the LPA discontinuity $y(\lambda^{+})\ne y(\lambda^{-})$ and the sticky limit.
This would provide an expression (not reported here) which is this sticky limit
of Eq.\ (\ref{sec:limit1}).

\section{Monte Carlo simulations and integral equation theory}
\label{sec:MC_IE}
In order to assess the reliability of the LPA, we will compare in Sec.\ \ref{sec:results}
with specialized MC simulations. In addition, prompted by the
results of Ref.\ \onlinecite{Santos08}, we will also compare LPA with standard integral
equation theories, such as PY and HNC.\cite{Hansen86}
\subsection{Monte Carlo simulations}
\label{subsec:MC}

We have employed the conventional Monte Carlo simulation on an NVT ensemble with periodic boundary
conditions which in one dimension means that the system {is} treated as a ring. $N = 5\times 10^4$
penetrable-rod particles were displaced according to the Metropolis algorithm
to create an initial sample of configurations.
Following the equilibration stage, each run
{is} divided into 20 basic simulation blocks, in which $10^5$ measurements {are} performed
to collect correlation functions data. $100$ trial moves per particle {are implemented} between each measurement,
so that $10^{13}$ equilibrium configurations {are} generated in each run.

In order to speed up the simulation process the particles are labeled such that they create
a consecutive sequence in a clockwise order.
Calculation of a potential of a particle $i$ in a given configuration then reduces to a searching
for the highest label $j>i$ and the lowest label $k<i$ associated with the particles still interacting
with the particle $i$. In contrast to the case of impenetrable spheres in one dimension
the order of particles changes so that a relabeling must be undertaken after each shift
 of a particle. Obviously, at {higher} temperatures the number of penetration can be high,
 which makes the calculations more demanding compared to hard body systems.

There are in general two routes for the evaluation of the pressure.
Determination of the pressure using a mechanical (virial) route relies
on an ensemble average of a virial, i.e. a quantity involving the forces acting
on all the particles. Alternatively, a thermodynamic expression relates pressure
to the volume derivative of the free energy and is implemented by calculating
\an{the free energy} change associated with small virtual change of volume.
However, for \an{systems} with discontinuous interaction both mechanical and thermodynamic
approaches become identical. Specifically, for the PSW fluid model both approaches reduce
on a calculation of distribution function at $r=1$ and $r=\lambda$ \an{[see Eq.\ \eqref{sec:eqs2}]}.

\subsection{Integral equations}
\label{subsec:ie}
The presence of penetrability does not pose any particular difficulties
to standard integral equation theories. As a matter of fact these have been
already employed in the PS case\cite{Malijevsky06} and in the PSW case\cite{Santos08}
within standard approximations where the one-dimensional Ornstein--Zernike  equation
\begin{eqnarray}
\label{sec:ie1}
h\left(r\right)=c(r)+\rho\int_{-\infty}^\infty \,dr' c\left(|r-r'|\right) h\left(r'\right)~, 
\end{eqnarray}
is associated with a PY closure
\begin{eqnarray}
c\left(r\right)=f\left(r\right) y\left(r\right)~,
\label{sec:ie2}
\end{eqnarray}
or with \an{an HNC closure}
\begin{eqnarray}
c\left(r\right)=f\left(r\right)y\left(r\right)+y\left(r\right)-1-\ln y\left(r\right)~.
\label{sec:ie3}
\end{eqnarray}
We have solved the PY and HNC integral equations using a Zerah's
algorithm \cite{Zerah85} with {up to} $2^{12}$ grid points depending {on the considered state point.}

\section{Results within the LPA}
\label{sec:results}
In this Section we compare numerical results stemming from the LPA with
MC simulations and integral equation theories (PY and HNC) for
both RDF (where we will consider the improved mLPA) and equation of state (at the level of the simple LPA).
\subsection{Results for  $g(r)$}
\label{subsec:g(r)}
As a first approach to assess the performance of the LPA, we consider the RDF
$g(r)$ for two representative state points.

The well is kept fixed at
$\Delta/\sigma=0.5$ and temperature is also fixed by the attractive energy scale
so that $k_B T/\epsilon_a=1$. Figure \ref{fig:fig3} depicts the
behavior of  $g(r)$ for a density $\rho\sigma=0.2$ and an energy ratio $\epsilon_r/\epsilon_a=5$,
which is well above the stability threshold value $\epsilon_r/\epsilon_a=2$.\cite{Santos08}
The stability threshold is then probed in Fig.\ \ref{fig:fig4}, whereas a higher
density  $\rho \sigma=0.8$
is tested in Fig.\ \ref{fig:fig5} with all other parameters identical to those of Fig.\ \ref{fig:fig3}.

In all cases, mLPA results (that only differ from the LPA ones within the interaction range, $0<r<\lambda$) are compared with MC simulations
and integral equations and follow the expected trend. For low densities ($\rho \sigma=0.2$) and low penetrability
($\epsilon_r/\epsilon_a=5$) mLPA, PY, and HNC all provide very accurate descriptions of MC data
with a very tiny difference in the well region $1\le r/\sigma\le 1.5$, where the integral equations predict a slight curvature of $g(r)$,
while the mLPA confirms the practically linear shape of the simulation data. Moreover, a blow-up of $g(r)$ in the deep core region  ($0\leq r\leq \Delta$) shows that the mLPA is very accurate, while the PY and HNC theories underestimate and overestimate, respectively, the MC data. The same good performance of the mLPA is also observed for a much larger penetrability ($\epsilon_r/\epsilon_a=2$), provided the density is relatively low ($\rho\sigma=0.2$), as shown in Fig.\ \ref{fig:fig4}. This is consistent with Fig.\ \ref{fig:fig2}, according to which the density $\rho\sigma=0.2$ lies in the region where the LPA is expected to be accurate for any temperature when  $\epsilon_r/\epsilon_a=2$ and $\Delta/\sigma=0.5$. As for the integral equations, they are also rather accurate for the case considered in Fig.\ \ref{fig:fig3}, although they still show a slight curvature inside the well and slightly deviate from the MC results for $r<\Delta$. Differences begin to
be relevant at high-density ($\rho\sigma=0.8$), mostly inside the core
$0<r/\sigma<1$ and in the contact values $r=\sigma^{+}$. Again, this agrees with Fig.\ \ref{fig:fig2}, which  shows that the state $(\rho\sigma,k_BT/\epsilon_a)=(0.8,1)$ is practically on the boundary line corresponding to $\epsilon_r/\epsilon_a=5$ and $\Delta/\sigma=0.5$. In any case, Fig.\ \ref{fig:fig5} shows that the best general agreement with the MC results is presented by the mLPA, followed by the HNC theory, which, however, predicts reasonably well the peaks of $g(r)$, but not the minima.

We have explicitly checked (not shown) that for smaller values of the well width $\Delta$, PSW results increasingly
tend to the SPS counterpart, as anticipated.

\subsection{Results for equation of state}
\label{subsec:eos}
Next we turn to the analysis of thermodynamics within LPA. As anticipated
(see Section \ref{sec:eos}), the lack of an exact solution gives rise to
thermodynamical inconsistencies where compressibility, virial, and energy routes
all give rise to different results. The consistency degree among different routes
is a (partial) signature of the LPA performance, as discussed in Sec.\ \ref{sec:limits}.
In Fig.\ \ref{fig:fig6} we report the behavior of $\beta p$ as a function of the
reduced density $\rho \sigma$. Once again, we fix the width of the
well $\Delta=0.5 \sigma$ and the energy ratio $\epsilon_r/\epsilon_a=5$, and
consider two different temperatures $k_B T/\epsilon_a=1$ {(top panel)}
and  $k_B T/\epsilon_a=5$ {(bottom panel)}. In the former case
different routes give practically indistinguishable results up to $\rho\sigma\approx 0.8$, whereas
in the latter a difference is clearly visible at densities higher than
$\rho \sigma \approx 0.6$ with energy, virial, and compressibility routes
having decreasing $\beta p$ for identical values of $\rho \sigma$.
Similar results are observed at the stability edge $\epsilon_r/\epsilon_a=2$,
as shown in Fig.\ \ref{fig:fig7}. We remark that higher temperatures effectively
correspond to higher penetrability, as particles have relatively
more attractive energies, as compared to the positive repulsive barrier,
and hence they can compenetrate more. Therefore pressure differences
among different thermodynamical routes can be reckoned as a rough measure
of the breakdown of LPA. On the other hand, consistency
among different routes does not necessarily means ``exact'' results,
as they can all converge to the incorrect value.

A comparison with MC numerical simulations is therefore also included
in Figs.\  \ref{fig:fig6} and \ref{fig:fig7}.
Somewhat surprisingly, this suggests that the
virial route is the closest to the true value for the pressure,
with both compressibility and energy routes always lying on the opposite side
with the latter \an{being} the farthest from the MC results.


In order to compare with LPA, we have carefully scanned a wide range of temperatures
and densities within the region $0 \le \rho \sigma \le 1$ where LPA provides
consistent thermodynamics as remarked.
Within this region we found no signature of fluid-fluid transition line as expected.
Our preliminary numerical results for higher densities, where strong overlapping among different
particles is enforced, provide a clear evidence of phase separation.
As the main emphasis of the present paper is on analytical approximations, this point will
be discussed in some detail elsewhere.


\subsection{Results for Fisher-Widom line}
\label{subsec:fw}
Let us follow the recipe given in Sec.\  \ref{sec:fw} to locate the
FW line. In Fig.\ \ref{fig:fig8} we report the quantities
$p\sigma/\epsilon_a$ and $\rho \sigma$ as a function of $k_B T/\epsilon_a$
for $\Delta=\sigma$ and decreasing values of the ratio $\epsilon_r/\epsilon_a$.
The case $\epsilon_r/\epsilon_a \to \infty$ is the one addressed in the original
FW work on the one-dimensional SW fluid.\cite{Fisher69}
We remind that above the FW line, $g(r)-1$ has oscillatory behavior, whereas it is
exponentially decaying below it,
and it is located in the  homogeneous fluid region of the phase diagram,
above the critical temperature if phase separation is present. 


As the repulsive barrier becomes finite, the region of monotonic behavior increases
for large $k_B T/\epsilon_a$ whereas it remains essentially unchanged for lower
temperatures. This is not surprising as penetrability (\emph{i.e.}, finite repulsive barrier)
favors the onset of a critical region. Somewhat more surprising is the
fact that this happens in the high- rather than in the low-temperature region.
A similar feature is also appearing in the $\rho$-$T$ plane {(see bottom panel)}.
In order to test the effect of different
width values, we have repeated the same calculation for $\Delta=0.5\sigma$.
Results are presented in Fig.\ \ref{fig:fig9} and are in agreement (in the limit $\epsilon_r/\epsilon_a \to \infty$) with
results for the one-dimensional SW fluid presented in Ref.\ \onlinecite{Perry72}
for a hard-core to well-width ratio equal to $2$ (see Fig.\ 1 in Ref.\ \onlinecite{Perry72}). For this well width the influence
of the ratio $\epsilon_r/\epsilon_a$ on the FW line is much less important.

Although we have been unable to find a simple physical explanation for this
behavior, we remark that the sensitivity of the FW line to the barrier height occurs as the density decreases. Consider for instance the density $\rho\sigma=0.1$ for models with $\Delta=\sigma$. In the SW case ($\epsilon_r/\epsilon_a\to\infty$) the decay of the RDF  changes from monotonic to oscillatory as one increases the temperature and crosses the value $k_BT/\epsilon_a\simeq 2.2$.
In the case of the PSW model with $\epsilon_r/\epsilon_a=5$,  according to the LPA, the transition takes place at $k_BT/\epsilon_a\simeq 2.8$. If the density is sufficiently low ($\rho\sigma\lesssim 0.076$ for $\epsilon_r/\epsilon_a=5$), the \an{asymptotic decay of $g(r)-1$} is monotonic for any temperature, while this effect is absent in the impenetrable SW limit. One might argue that this influence of the energy ratio $\epsilon_r/\epsilon_a$ on the high-temperature branch of the FW line is an artifact of the LPA since the latter approximation is \emph{a priori} restricted to low temperatures. On the other hand, this high-temperature branch also corresponds to low densities,  counterbalancing the penetrability effect and making the LPA presumably accurate. As a matter of fact, the FW lines plotted in the {top panels} of Figs.\ \ref{fig:fig8} and \ref{fig:fig9} are obtained from the three thermodynamic routes but the three curves are, in each case, indistinguishable each oth!
 er. In other words, the FW lines are well inside the regions in Fig.\ \ref{fig:fig2} where the LPA is thermodynamically consistent from a practical point of view.

\section{Conclusions {and outlook}}
\label{sec:conc}
One-dimensional fluids with nearest-neighbor interactions admit
an exact analytical solution for both structural and thermophysical
properties with a well defined protocol.\cite{Salsburg53,Heying04}
Nearest-neighbor interactions, {in turn}, require
a well defined hard-core term in the pair-wise potential preventing
superpositions and particle exchanges which is the crucial ingredient
necessary for the exact solution. The absence of the above constraint, on the other hand,
allows the presence of critical phase transitions, in spite of the
one-dimensional character of the system, which are fully absent
in the hard-core counterparts.

Effective pair interactions with a soft-repulsive component are
well-known features of polymer solutions and colloidal suspensions.\cite{Likos01,Barrat03}
Among many different model potentials,\cite{Likos01} with various degrees of core softness,
penetrable spheres (PS) stands out for its simplicity.\cite{Malijevsky06}
In this model, the infinite repulsive energy is reduced to a finite one,
thus introducing an effective ``temperature'' into an otherwise
athermal hard-sphere system. This potential model lacks of attractive
interactions but these can be accounted for in the penetrable square-well
(PSW) companion model where an attractive short-range square-well is added
to the PS model.\cite{Santos08}

At sufficiently low temperatures, thermal energy cannot overcome the repulsive barrier and penetrability is low, whereas
at high temperatures different particles can interpenetrate to a significant extent.
Hence, within this framework, low- and high-temperature and low- and high-penetrability
terminology can be used synonymously.

In this work we have studied structural and thermodynamic properties of the PSW model.
Using a low-penetrability approximation (LPA) akin to that discussed for PS,\cite{Malijevsky06} we
have considered rather interesting issues specific of the presence of attractive interactions
(and thus absent in the PS model) such as fluid-fluid phase separation or
the existence of a Fisher-Widom line.\cite{Fisher69} This is a pseudo-transition
associated with a {clear-cut} change, from oscillatory to monotonic, in the asymptotic decay properties of the
radial distribution function, as transition line is approached, even in those cases   where the existence of a critical region is prevented by
rigorous theorems (e.g.the SW one-dimensional fluid). It requires the simultaneous
presence of attractive and repulsive energies and hence
it cannot exist for the simpler PS model

Our LPA has been devised to reduce to that of PS in the limit of no well.
We have assessed its performance by comparing it with exact results\cite{Santos08} in the low-density limit and by comparing with MC simulations
and PY and HNC integral equation theories for larger densities where exact analytical results
do not exist. We found that it reproduces a significant portion of the
$T$-$p$ parameter space at the level of pair correlation function, the main difference being in the
{penetrability region $0<r<\sigma$}. At \an{odds with} its square-well counterpart, PSW \an{thermodynamics} depends
upon the chosen route in view of the inconsistencies introduced by the LPA.
We have quantified the inconsistencies among virial, compressibility, and energy routes
and discussed how they reflect into the computation of the Fisher--Widom line.
In all considered cases, we found a magnification at large temperatures of the monotonic regime region
as penetrability increases and a much smaller, if any,
modification, at lower temperatures.
{In all cases the FW line \an{is} found within the region where LPA is expected to be accurate
as thermodynamic inconsistencies are small. Within the density region $0\le \rho \sigma \le 1$,
we have found no sign of a fluid-fluid phase separation, although both fluid-fluid
and fluid-solid transitions are expected at higher densities.}

In the limit of infinitely narrow and deep well, PSW has been shown
to reduce to a penetrable version of Baxter adhesive model,\cite{Baxter68}
which violates the stability condition set for a well defined thermodynamic
limit.\cite{Santos08}

As the main weaknesses of LPA for the PSW
stems mainly from a non-adequate representation of the penetrable
region $0<r/\sigma<1$, we have then discussed how a simple
modification of the radial distribution function in this region
gives a significant improvement when tested against MC results
under rather demanding conditions.

This paper is part of an on-going effort on PSW outlined
in our previous work.\cite{Santos08} Future work
will address a complementary approximation (the \an{high-penetrability} limit)
and its matching with the LPA discussed in the present
paper, so that the entire parameter $T$-$p$-$\rho$ space can be
discussed with some comfortable degree of confidence. This will resolve some of the
subtle points with no conclusive answer left by the present paper.
{In addition, a detailed investigation of the high density region $\rho \sigma >1$ is
underway and will be reported elsewhere.}.

\begin{acknowledgments}
The work of R.F. and A.G. was supported by the Italian MIUR through a grant PRIN-COFIN 2007B57EAB
(2008/2009). A.M. is grateful to the support of The Ministry of Education, Youth,
and Sports of the Czech Republic, under Project No. LC512 (Center
for Biomolecules and Complex Molecular Systems). The research of A.S. was supported by the Ministerio de Educaci\'on y Ciencia (Spain) through Grant No.\
FIS2007-60977 (partially financed by FEDER funds) and by the Junta
de Extremadura through Grant No.\ GRU09038.
\end{acknowledgments}

\appendix
\section{Analysis of the continuity of $y(r)$ within LPA}
\label{app:appa}
{}From Eq.\ (\ref{sec:low7}) we have that if $r<2$,
\begin{eqnarray}
\rho g(r)=\frac{h\zeta}{1-q} e^{-\zeta r}+
\begin{cases}
0,&0\leq r<1,\\
\psi_{1}(r-1),&1<r<1+\Delta,\\
\psi_{1}(r-1)-q\psi_{1}(r-1-\Delta),&1+\Delta<r<2.
\end{cases}
\label{app:a1}
\end{eqnarray}
The explicit expressions of $\psi_{1}(r)$ is, from Eq.\ (\ref{sec:low8}),
\begin{eqnarray}
\label{app:a2}
\psi_{1}\left(r\right)&=&\frac{\zeta}{1-q} e^{-\zeta r}\left(1+\frac{h\zeta}{1-q} r\right).
\end{eqnarray}

The continuity condition of $y(r)$ at $r=1$ is then given by condition
\begin{eqnarray}
\label{app:a3}
\frac{1}{1-\gamma_r}\frac{h\zeta}{1-q}e^{-\zeta}=\frac{1}{1+\gamma\gamma_r}\left[\frac{h\zeta}{1-q}e^{-\zeta}+\psi_{1}(0)\right],
\end{eqnarray}
which is identically satisfied, so that
\begin{eqnarray}
\label{app:a4}
\rho y\left(1\right)&=&\frac{\zeta}{\left(1-q\right)\gamma_r\left(1+\gamma\right)}.
\end{eqnarray}
However, $y(r)$ is discontinuous at $r=\lambda=1+\Delta$:
\begin{eqnarray}
\rho y\left(\lambda^-\right)&=&\frac{1}{1+\gamma\gamma_r}\left[\frac{h\zeta}{1-q}e^{-\zeta\lambda}+\psi_{1}(\Delta)\right]  \nonumber \\
\label{app:a5}
&=&\frac{\zeta q}{\gamma_r\gamma(1-q)}\left[1+\frac{\gamma_r(1+\gamma)}{1+\gamma\gamma_r}\frac{h\zeta\Delta}{1-q}\right],
\\
\rho y\left(\lambda^+\right)&=&\frac{h\zeta}{1-q}e^{-\zeta\lambda}+\psi_{1}(\Delta)-q\psi_1(0) \nonumber \\
&=&\frac{\zeta q}{\gamma_r\gamma(1-q)}\left[1+{\gamma_r(1+\gamma)}\frac{h\zeta\Delta}{1-q}\right].
\label{app:a6}
\end{eqnarray}
The jump is then given by
\begin{eqnarray}
\label{app:a7}
\rho \left[y(\lambda^+)-y(\lambda^-)\right]=\frac{\gamma_r(1+\gamma)}{1+\gamma\gamma_r}\frac{hq\zeta^2\Delta}{(1-q)^2},
\end{eqnarray}
and the value used as an estimate of the point is then given by the average of the left and right limits
\begin{eqnarray}
\label{app:a8}
{\rho} \frac{y\left(\lambda^+\right)+y\left(\lambda^-\right)}{2}=
\frac{\zeta q}{\gamma_r\gamma\left(1-q\right)}\left[1+{\gamma_r\left(1+\gamma\right)}
\frac{1+\gamma\gamma_r/2}{1+\gamma\gamma_r}\frac{h\zeta\Delta}{1-q}\right].
\end{eqnarray}
\section{The sticky-penetrable-sphere (SPS) model}
\label{app:appb}
In this Appendix, we provide a connection with the SPS introduced in Ref.\ \onlinecite{Santos08}. This is the penetrable analogue
of Baxter's sticky-hard-sphere (SHS) well known model.\cite{Baxter68}
The SPS limit can be obtained
by considering the limit $\Delta \to 0$ and $\epsilon_a \to \infty$ so that
$\alpha=\gamma \Delta$ remains finite, hence playing the role of an adhesivity
parameter. We then define SPS by the Mayer function\cite{Santos08}
\begin{eqnarray}
\label{app:b1}
f_{\text{SPS}}\left(r\right)&=&\gamma_rf_{\text{SHS}}\left(r\right)~,
\end{eqnarray}
where
\begin{eqnarray}
\label{app:b2}
f_{\text{SHS}}\left(r\right)&=&f_{\text{HS}}\left(r\right)+\alpha\delta_+\left(r-\sigma\right)
\end{eqnarray}
is the Mayer functions of the SHS potential and
\begin{eqnarray}
\label{app:b3}
\delta_+\left(r\right)\equiv\lim_{a\to 0^+}\frac{\Theta\left(r\right)-\Theta\left(r-a\right)}{a}~.
\end{eqnarray}
The fluid parameters are then the
adhesivity coefficient $\alpha>0$, the penetrability
coefficient $\gamma_r$, and the density $\rho$.

{As} anticipated, the SPS fluid is thermodynamically unstable
in the sense discussed in Section \ref{sec:psw}. This can be seen
both because the required limit does not satisfy the sufficient
condition for stability $\epsilon_r > 2 \epsilon_a$,\cite{Santos08}
and directly using arguments akin to those used by Stell\cite{Stell91}
to prove the instability of the original Baxter's model\cite{Baxter68}
in dimensions greater than one.
Nonetheless it provides an overall consistency  testbench to
the performance of LPA within the well established
framework of SHS.

In the combined limit $\gamma\to\infty$ and $\Delta\to 0$ with
$\alpha=\gamma\Delta$, Eqs.\ (\ref{sec:low1}) and (\ref{sec:low1bis}) become
\begin{eqnarray}
\label{app:b4}
\widetilde{\Omega}^{\text{SPS}}(s) &=&\frac{1-\gamma_r}{s}+
\gamma_r\left(\alpha+\frac{1}{s}\right)e^{-s}~,
\end{eqnarray}
\begin{eqnarray}
\label{app:b5}
\widetilde{\Omega}_0^{\text{SPS}}(s)=\gamma_r\left(\alpha+\frac{1}{s}\right)e^{-s}~.
\end{eqnarray}
Using the first equality in Eq.\ (\ref{sec:low3}) it follows that
\begin{eqnarray}
\label{app:b6}
\rho&=&\frac{f/\zeta+1/\zeta+\alpha}{\alpha+1/\zeta+1/\zeta^2}~,
\end{eqnarray}
where
\begin{eqnarray}
\label{app:b7}
f&=&\frac{1-\gamma_r}{\gamma_r}e^{\zeta}~.
\end{eqnarray}

We then use the LPA  as given in Eq.\ (\ref{sec:low2})
to find
\begin{eqnarray}
\label{app:b8}
\rho\widetilde{G}^{\text{SPS}}\left(s\right)&=&\frac{f/\left(s+\zeta\right)+\left[\alpha+1/\left(s+\zeta\right)\right]e^{-s}}
{\left(\alpha+1/\zeta\right)-\left[\alpha+1/\left(s+\zeta\right)\right]e^{-s}}~,
\end{eqnarray}
whose inverse Laplace transform yields
the RDF,
\begin{eqnarray}
\label{app:b9}
\rho g^{\text{SPS}}\left(r\right)&=&\sum_{n=0}^\infty \psi_n^{\text{SPS}}\left(r-n\right)\Theta\left(r-n\right)~,
\end{eqnarray}
where
\begin{eqnarray}
\label{app:b10}
\psi_0^{\text{SPS}}(r)&=&\frac{f}{\alpha+1/\zeta}e^{-\zeta r}~,\\
\label{app:b11}
\psi_n^{\text{SPS}}(r)&=&\left(\frac{\alpha}{\alpha+1/\zeta}\right)^n\left[
\frac{f}{\alpha+1/\zeta}+
\sum_{k=1}^n\binom{n}{k}
\frac{1}{\alpha^{k}k!}\left(kr^{k-1}+\frac{f}{\alpha+1/\zeta}r^k\right)
+\delta(r)\right]e^{-\zeta r}~.
\end{eqnarray}

In the impenetrable limit $\gamma_r\to 1$ and $f\to 0$,
Eqs.\  (\ref{app:b6})--(\ref{app:b11}) reduce to the exact one-dimensional SHS counterparts,\cite{Tago75,Seaton86,Yuste93} as they should.

A word of caution is in order here.
Using Eqs.\ (\ref{app:b9})--(\ref{app:b11}), the cavity function $y(r)=g(r) e^{\beta \phi(r)}$
at contact $r=1$ is found to be discontinuous as
\begin{eqnarray}
\label{app:b12}
\rho y^{\text{SPS}}\left(1^{-}\right) &=& \frac{1}{\gamma_r} \frac{1}{\alpha + 1/\zeta}, \\
\label{app:b13}
\rho y^{\text{SPS}}\left(1^{+}\right) &=& \rho y^{\text{SPS}}\left(1^{-}\right) +
\frac{\alpha f}{\left(\alpha+1/\zeta\right)^2}.
\end{eqnarray}
Note that Eq.\ (\ref{app:b12}) is the sticky limit of the PSW value $\rho y(1)$, Eq.\ (\ref{app:a4}),
[recall that $y(r)$ is continuous at $r=1$ within the PSW] and is also the sticky limit
of the PSW value $\rho y(\lambda^{-})$, Eq.\ (\ref{app:a5}). On the other hand, Eq.\ (\ref{app:b13})
is the sticky limit of the PSW value $y(\lambda^{+})$, Eq.\ (\ref{app:a5}).
Therefore, the discontinuity of  $y^{\text{SPS}}(r)$ at $r=1$ is a direct consequence of the discontinuity of the PSW cavity function at $r=\lambda$.
Both discontinuities are artifacts of the LPA.
Again, this can be amended by an improved mLPA approach which \an{is} discussed in Sec.\ \ref{sec:limits}.

\section{Low-density expansion of the LPA}
\label{app:appc}

\an{Let us  compare} the LPA to order $\rho$ with the exact results. {}From Eqs.\ \eqref{sec:low3}--\eqref{sec:low5} we easily get
\begin{equation}
\zeta =\zeta_0 \rho+\zeta_1\rho^2+\mathcal{O}(\rho^3)
 \label{app:c1}
\end{equation}
 with
\begin{equation}
 \zeta_0=\gamma_r,\quad \zeta_1=\gamma_r^3(1-\gamma \Delta).
 \label{app:c2}
\end{equation}
Upon inserting the result into Eqs.\ (\ref{sec:low7}) and (\ref{sec:low8}), and after some algebra,
we find the correct zeroth order term $g_0(r)=g_0^{\text{exact}}(r)$ as given
in Eq.\ (\ref{subsec:comp2}), and
\begin{eqnarray}
\label{app:c3}
g_1\left(r\right)=\gamma_r^2\begin{cases}
(1-\gamma_r)\left[1-\gamma\frac{1+\gamma_r}{\gamma_r}\Delta-(r-1)\frac{1}{\gamma_r}\right],  & 0 \leq r< 1, \\
(1+\gamma\gamma_r)\left[1-\gamma\frac{1+\gamma_r}{\gamma_r}\Delta+(r-1)\frac{\gamma-\gamma_r-2\gamma\gamma_r}{\gamma_r(1+\gamma\gamma_r)}\right],  & 1< r< 1+\Delta, \\
2-2 \gamma\Delta-r,  & 1+\Delta< r\leq 2, \\
\gamma\left(2+\gamma\right)\left(r-2\right)-2\gamma\Delta, & 2\leq r\leq 2+\Delta,  \\
\left(2+2\Delta -r\right)\gamma^2, & 2+\Delta \leq r\leq 2+2\Delta,  \\
 0, & 2+2\Delta \leq r.
\end{cases}
\end{eqnarray}
Comparison between Eqs.\ \eqref{app:c3} and \eqref{subsec:comp3} shows that the LPA reproduces the exact result for $r\geq 1+\Delta$.
On the other hand, it fails to do so within the potential range. The differences between the first-order exact and LPA  cavity functions
are given by Eq.\ \eqref{subsec:comp4}.


\begin{thebibliography}{99}
\bibitem{Barrat03} J.-L. Barrat and J.-P. Hansen, \textit{Basic Concepts
for Simple and Complex Liquids} (Cambridge University Press, Cambridge, 2003).
\bibitem{Likos01} C.N. Likos, Phys. Rep. \textbf{348}, 267 (2001).
\bibitem{Watzlawek99} M. Watzlawek, C.N. Likos and H. L\"owen, Phys. Rev. Lett.
\textbf{82}, 5289 (1999)

\bibitem{Ballauff04} M. Ballauff and C.N. Likos, Angew. Chem. Int. Ed. \textbf{43},
2998 (2004)
\bibitem{Salsburg53} Z.W. Salsburg, R.W. Zwanzig, and J.G. Kirkwood, J. Chem.
Phys. \textbf{21}, 1098 (1953).
\bibitem{Fantoni03} See e.g. R. Fantoni, Ph.D. thesis, University of Trieste, 2003,
(unpublished), and references therein.
\bibitem{Malijevsky06} Al. Malijevsk\'y and A. Santos, J. Chem. Phys.
\textbf{124}, 074508 (2006).
\bibitem{Likos98} C.N. Likos, M. Watzalwek, and H. L\"owen, Phys. Rev. E
\textbf{58}, 3135 (1998).

\bibitem{Lang00} A. Lang, C.N. Likos, M. Watzlawek and H. L\"owen,
J. Phys: Cond. Mat. \textbf{12}, 5087 (2000)

\bibitem{Santos08} A. Santos, R. Fantoni, and A. Giacometti, Phys. Rev. E \textbf{77},
051206 (2008).
\bibitem{Ruelle69} D. Ruelle, \textit{Statistical Mechanics: Rigorous Results}
(Benjamin, London, 1969).
\bibitem{Fisher66} M.E. Fisher and D. Ruelle, J. Math. Phys. \textbf{7},
260 (1966).
\bibitem{Widom70} B. Widom and J.S. Rowlinson, J. Chem. Phys. \textbf{52}, 1670
(1970).
\bibitem{Torquato84} S. Torquato, J. Chem. Phys. \textbf{81}, 5079 (1984).
\bibitem{Rikvold85} P.A. Rikvold and G. Stell, J. Chem. Phys. \textbf{82}, 1014 (1985).
\bibitem{Stillinger76} F.H. Stillinger, J. Chem. Phys. \textbf{65}, 3968 (1976).
\bibitem{Louis00} A.A. Louis, P.G. Bolhuis, and J.-P. Hansen, Phys. Rev. E \textbf{62}, 7961 (2000).
\bibitem{Fisher69} M.E. Fisher and B. Widom, J. Chem. Phys. \textbf{50},
3756 (1969).
\bibitem{Corti98} D.S. Corti and P.G. Debenedetti, Phys. Rev. E \textbf{57},
4211 (1998).
\bibitem{Heying04} M. Heying and D.S. Corti, Fluid Phase Equilibria
\textbf{220}, 85 (2004).
\bibitem{Note2} It may be noted that
the  low-penetrability approximation presented here is identical to
 the low-temperature approximation  introduced in Ref.\ \protect\onlinecite{Malijevsky06}.
\bibitem{Yuste93} S.B. Yuste and A. Santos, J. Stat. Phys. \textbf{72},
703 (1993).
\bibitem{Santos06} A. Santos,  Mol. Phys. \textbf{104}, 3411 (2006).
\bibitem{Giacometti09} See e.g. A. Giacometti, G. Pastore, and F. Lado,
Mol. Phys. \textbf{107}, 555 (2009)


\bibitem{Hansen86} J.-P. Hansen and I.R. McDonald \textit{Theory of Simple Liquids}
(Academic Press, Amsterdam, 2006).


\bibitem{Evans93} R. Evans, J.R. Henderson, D.C.Hoye, A.O. Parry, and Z.A. Suber, Mol. Phys.
\textbf{80}, 755 (1993).
\bibitem{DeCarvalho94} R.J.F. Leote De Carvalho, R. Evans, D.C. Hoyle, and J.R. Henderson,
J. Phys.: Condens. Matter \textbf{6}, 9275 (1994)
\bibitem{Vega95} C. Vega, L.F. Rull, and S. Lago, Phys. Rev. E \textbf{51}, 3146 (1995).
\bibitem{Dijkstra00} M. Dijkstra and R. Evans, J. Chem. Phys. \textbf{112}, 1449 (2000).
\bibitem{Tarazona03} P. Tarazona, E. Chac\'on, and E. Velasco, Mol. Phys. \textbf{101}, 1595.
(2003).


\bibitem{Note3}
It is interesting to remark that $\lim_{\beta\to\infty}\zeta(\rho,\beta)= 0$ if $\rho<(1+\Delta/2)^{-1}$,
while $\lim_{\beta\to\infty}\zeta(\rho,\beta)=\zeta_0(\rho)\neq 0$
if $\rho>(1+\Delta/2)^{-1}$, where $\zeta_0(\rho)$ is the solution of $\rho^{-1}=\zeta_0^{-1}+(1-\lambda e^{-\zeta_0\Delta})/(1-e^{-\zeta_0\Delta})$.



\bibitem{Zerah85} G. Zerah, J. Comput. Phys. \textbf{61}, 280 (1985).
\bibitem{Perry72} P. Perry and S. Fisk, J. Chem. Phys. \textbf{57}, 4065 (1972).
\bibitem{Baxter68} R.J. Baxter, J. Chem. Phys. \textbf{49}, 2770 (1968).
\bibitem{Stell91} G. Stell, J. Stat. Phys. \textbf{63}, 1203 (1991).
\bibitem{Tago75} Y. Tago and S. Katsura, Can. J. Phys. \textbf{53}, 2587 (1975).
\bibitem{Seaton86} N.A. Seaton and E.D. Glandt, J. Chem. Phys. \textbf{84},
4595 (1986).
\end{thebibliography}

\clearpage
\centerline{\bf  Figure captions}

\begin{figure}[h]
\caption{(Color online) The PSW potential (top panel). The middle and bottom panels sketch the
different behavior of the SW and  PSW models, respectively. In the SW case there
exists a hard core (black inner sphere) and an interaction range
(light blue outer sphere) so two spheres on a line can either non interact
(A) or attract each other as the corresponding interaction spheres overlap (B).
As a consequence, different spheres cannot interchange positions on a
one-dimensional line and the problem is analytically solvable.
In the PSW the core is soft (red inner sphere) and hence we can have
in addition
to configurations (A) and (B) identical to the SW case, also the case
where the internal cores overlap such as (C) and (D). Different spheres
can then interchange position
and the problem is a many-body one.
\label{fig:fig1}}
\end{figure}

\begin{figure}[h]
\caption{(Color online) Schematic phase diagram in the $\rho\sigma$-$k_B
  T/\epsilon_a$ space showing
the region where LPA can be considered as reliable. The curves
correspond, from top to bottom, to the cases
$(\epsilon_r/\epsilon_a,\Delta)=(5,0.5)$, $(5,1)$, and $(2,0.5)$. The points below each curve represent states
where the relative deviation between the three routes to the pressure is smaller than 5\%.
\label{fig:fig2}}
\end{figure}

\begin{figure}[h]
\caption{(Color online) Results for the radial distribution function $g(r)$ versus
$r/\sigma$ with  $\Delta/\sigma=0.5$, $k_B T/\epsilon_a=1$,
  $\epsilon_r/\epsilon_a=5$, and $\rho \sigma=0.2$. Predictions from the modified LPA given by Eq.\ \protect\eqref{sec:limit1}
(long dashed line) are compared with both MC results (solid line) and
PY and HNC integral equations (short dashed and dotted lines,
respectively). In the inset a magnification of the $r<\sigma$ region
is shown. 
\label{fig:fig3}}
\end{figure}

\begin{figure}[h]
\caption{(Color online) Same as in Fig.\ \ref{fig:fig3} at the instability threshold
$\epsilon_r/\epsilon_a=2$. All other parameters are as in Fig.\ \ref{fig:fig3}.
\label{fig:fig4}}
\end{figure}

\begin{figure}[h]
\caption{(Color online) Same as in Fig.\ \ref{fig:fig3} at a higher density
$\rho \sigma=0.8$. All other parameters are as in Fig.\ \ref{fig:fig3}.
\label{fig:fig5}}
\end{figure}

\begin{figure}[h]
\caption{(Color online) Plot of $\beta p\sigma$ vs $\rho \sigma$ for $\Delta/\sigma=0.5$,
$\epsilon_r/\epsilon_a=5$, and $k_B T/\epsilon_a=1$ {(top panel)}
and $k_B T/\epsilon_a=5$ {(bottom panel)}. Different curves
refer to different routes. \an{The symbols denote MC simulation results}.
\label{fig:fig6}}
\end{figure}

\begin{figure}[h]
\caption{(Color online) Same as in Fig.\ \ref{fig:fig6}, except that $\epsilon_r/\epsilon_a=2$.
\label{fig:fig7}}
\end{figure}

\begin{figure}[h]
\caption{(Color online) Plot of the Fisher--Widom transition line in the $p\sigma/\epsilon_a$
vs $k_B T/\epsilon_a$ plane {(top panel)} and in the $\rho
\sigma$ vs $k_B T/\epsilon_a$ plane {(bottom
panel)}. Here $\Delta/\sigma=1$ and  $\epsilon_r/\epsilon_a=(\infty, 10,5)$. Note that, except in the
SW case ($\epsilon_r/\epsilon_a = \infty$), $\zeta/\sigma \ne \beta
p$. Note also that in these cases the three routes to the pressure
are not distinguishable one from the other on the graph scale.
\label{fig:fig8}}
\end{figure}

\begin{figure}[h]
\caption{(Color online) Same as in Fig.\ \ref{fig:fig8}, except that $\Delta/\sigma=0.5$.
\label{fig:fig9}}
\end{figure}

\clearpage

%
%
\begin{figure}[tbp]
\includegraphics[width=.5 \columnwidth]{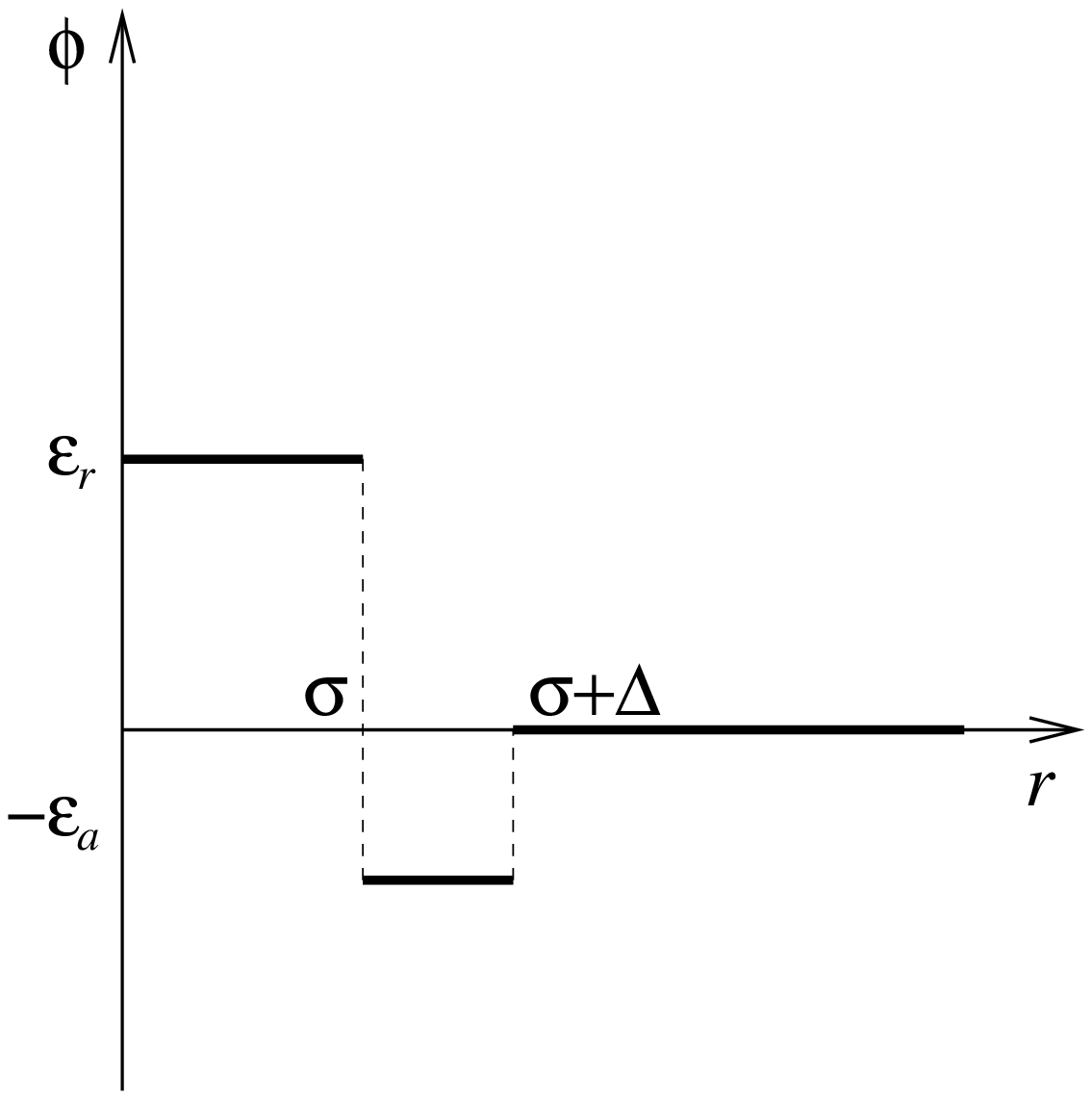} \\
\includegraphics[width=.5 \columnwidth]{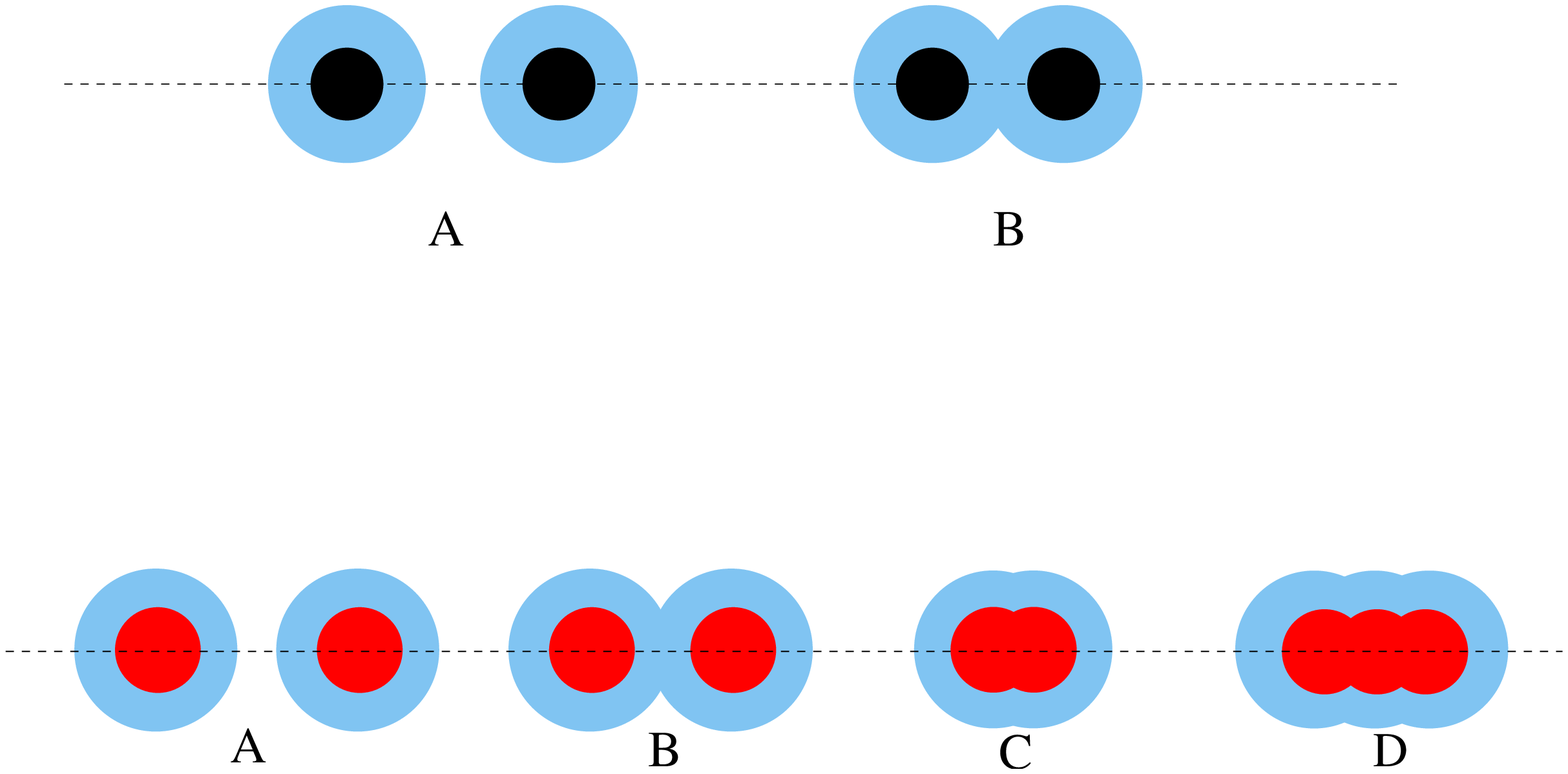} \\
\end{figure}
\begin{center}
FIG.\ \ref{fig:fig1}. Fantoni \textit{et al.} (JCP)
\end{center}

\clearpage

%
%
\begin{figure}[tbp]
\includegraphics[width=14cm]{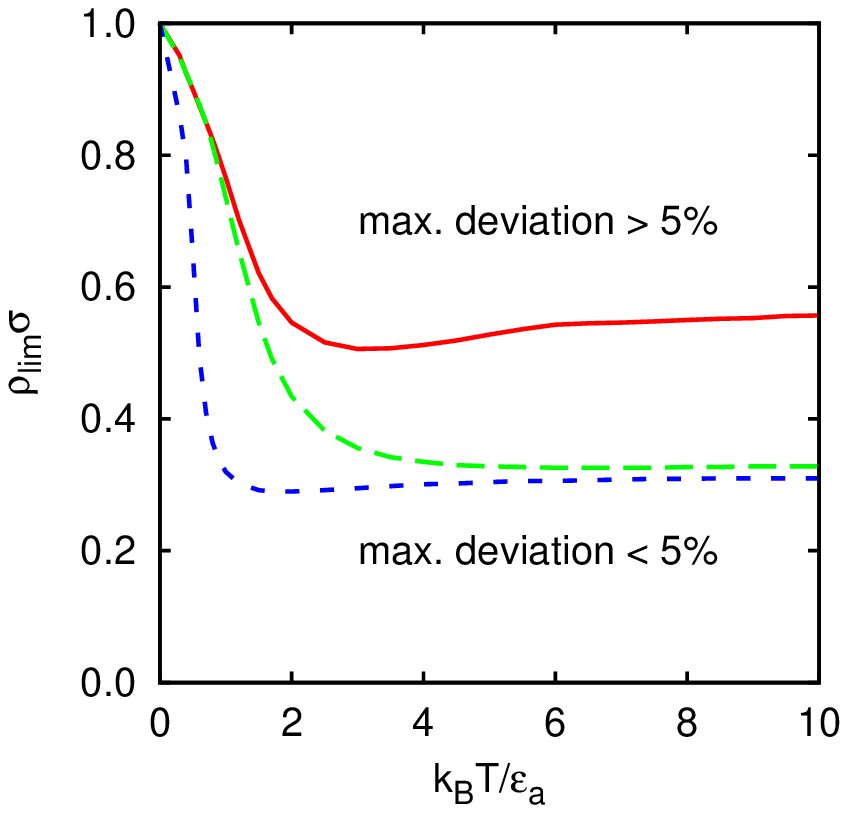}
\end{figure}
\begin{center}
FIG.\ \ref{fig:fig2}. Fantoni \textit{et al.} (JCP)
\end{center}

\clearpage

%
%
\begin{figure}[tbp]
\includegraphics[width=14cm]{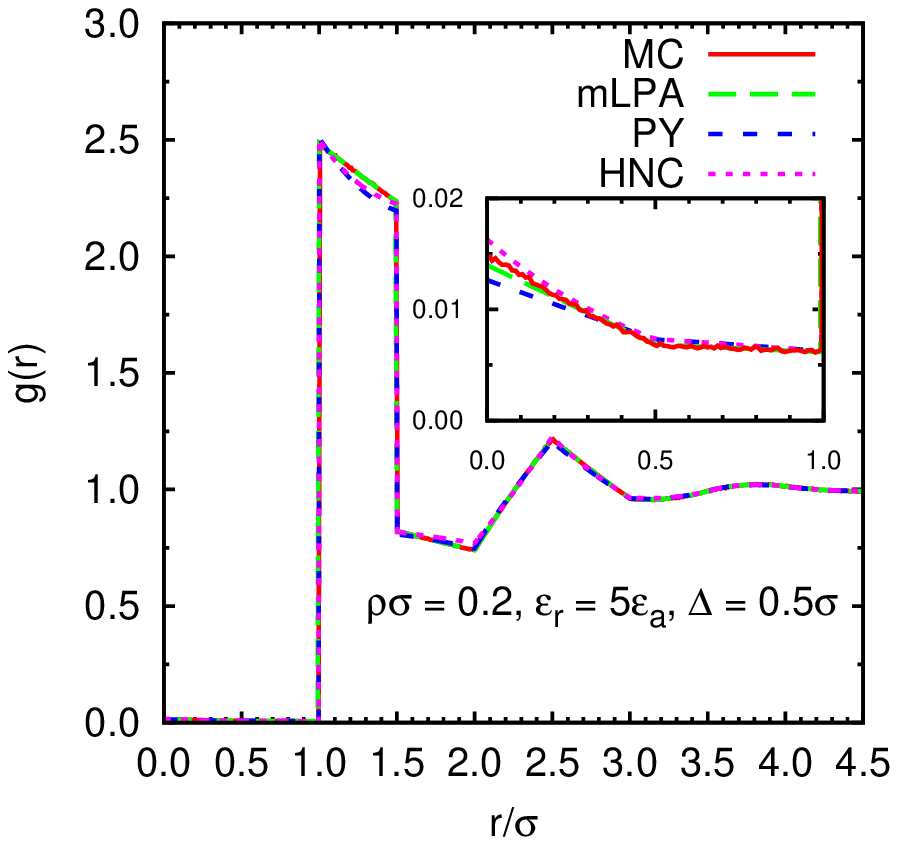}
\end{figure}
\begin{center}
FIG.\ \ref{fig:fig3}. Fantoni \textit{et al.} (JCP)
\end{center}

\clearpage

%
%
\begin{figure}[tbp]
\includegraphics[width=14cm]{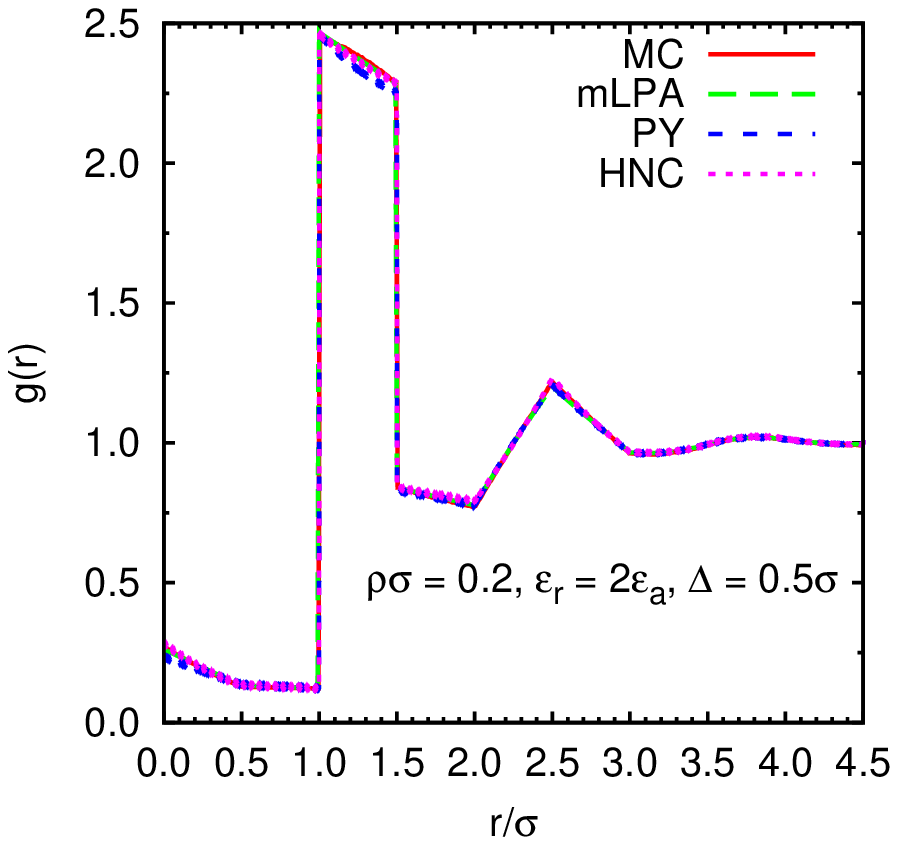}
\end{figure}
\begin{center}
FIG.\ \ref{fig:fig4}. Fantoni \textit{et al.} (JCP)
\end{center}

\clearpage

%
%
\begin{figure}[tbp]
\includegraphics[width=14cm]{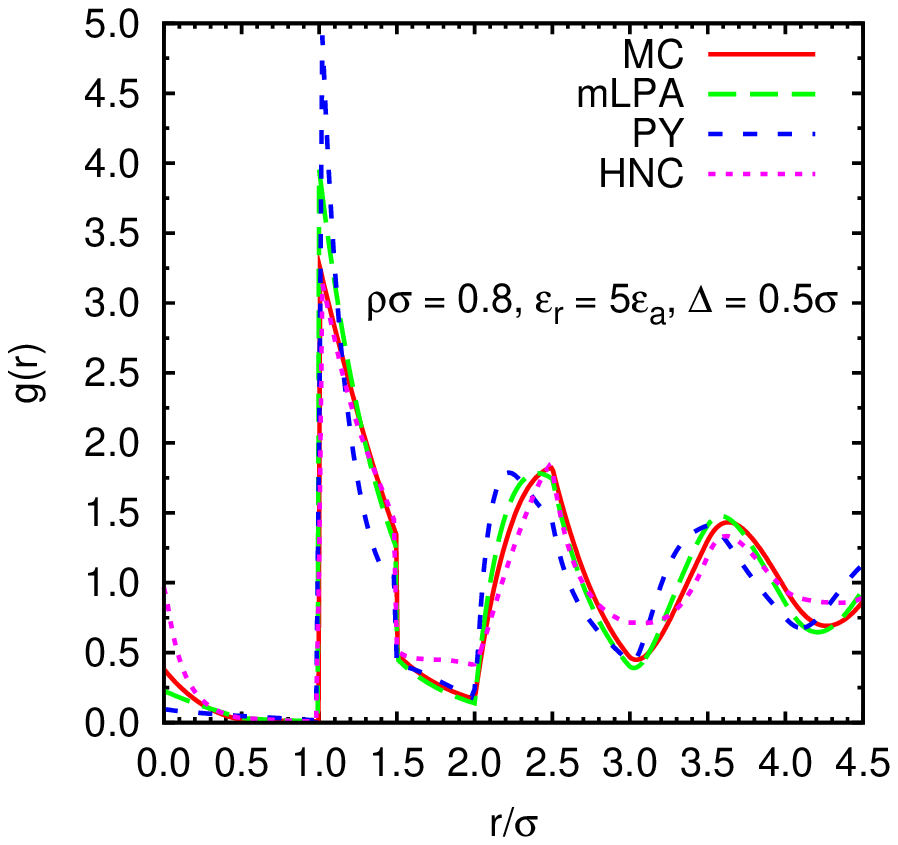}
\end{figure}
\begin{center}
FIG.\ \ref{fig:fig5}. Fantoni \textit{et al.} (JCP)
\end{center}

\clearpage

%
%
\begin{figure}[tbp]
\includegraphics[width=14cm]{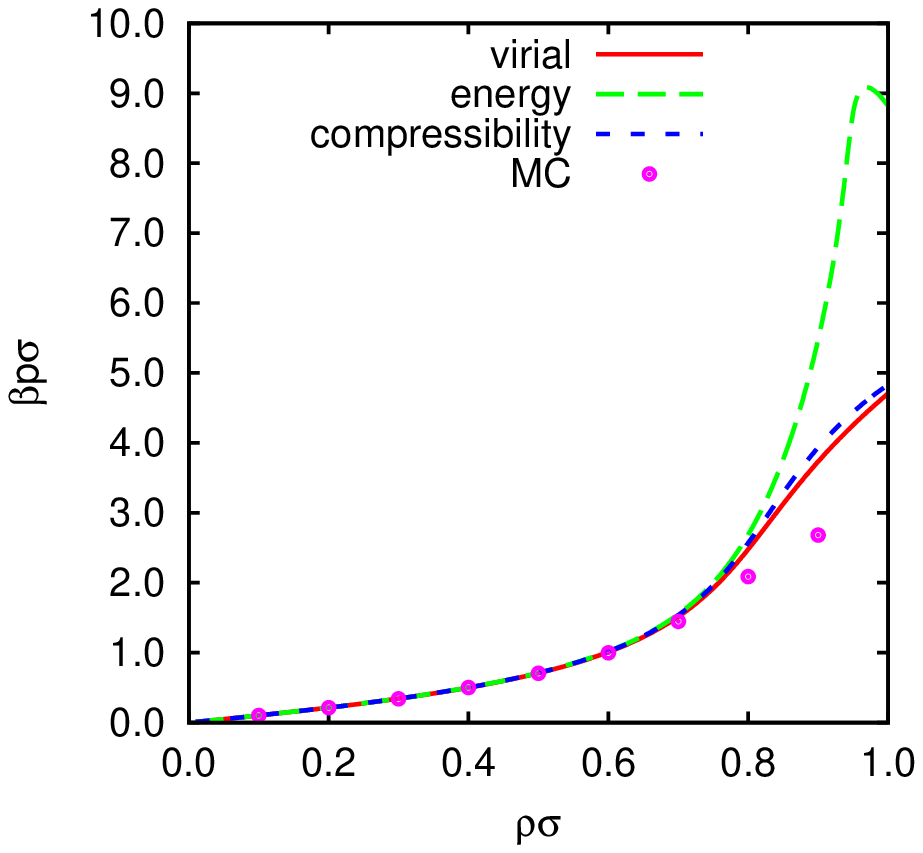}
\includegraphics[width=14cm]{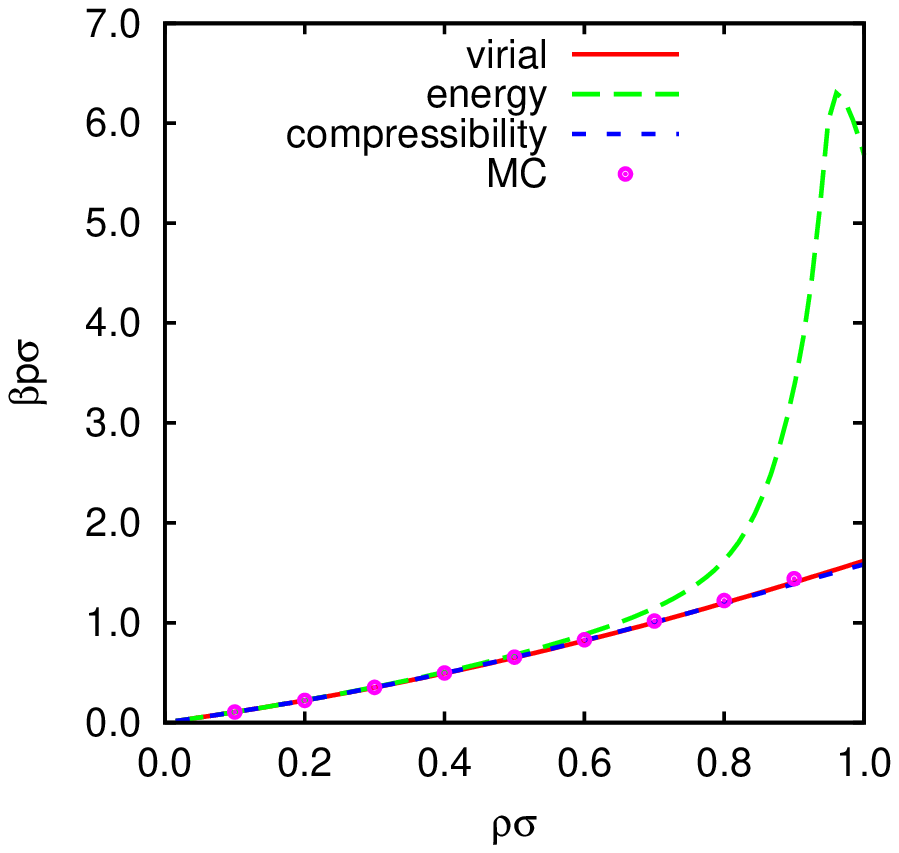}
\end{figure}
\begin{center}
FIG.\ \ref{fig:fig6}. Fantoni \textit{et al.} (JCP)
\end{center}

\clearpage

%
%
\begin{figure}[tbp]
\includegraphics[width=14cm]{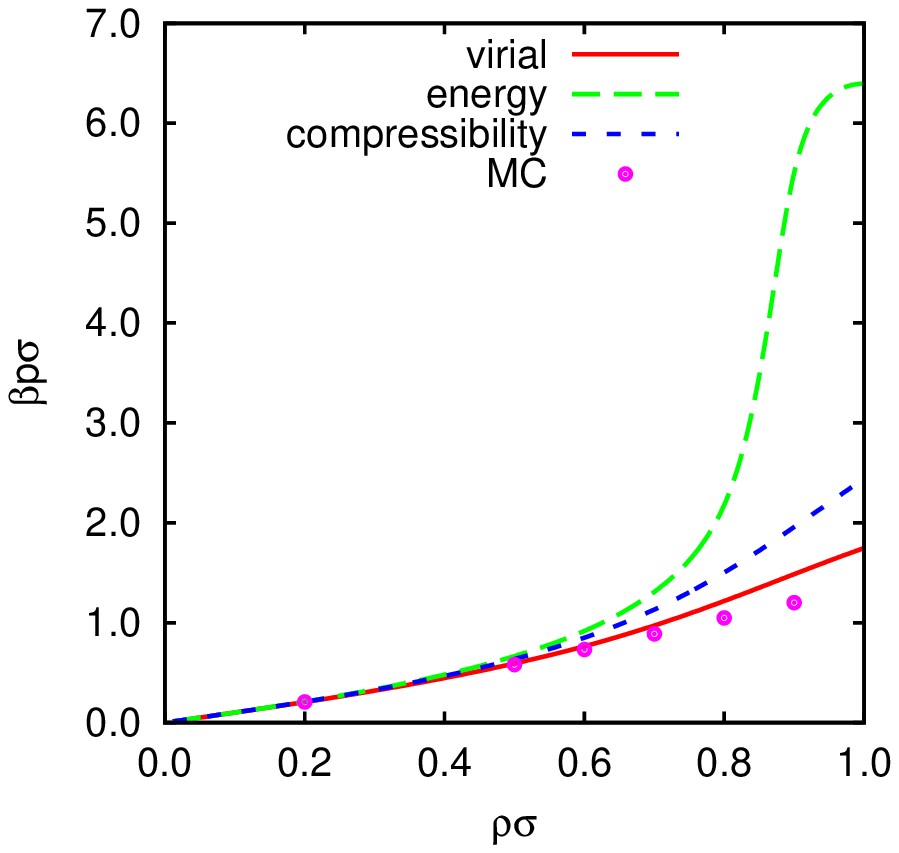}
\includegraphics[width=14cm]{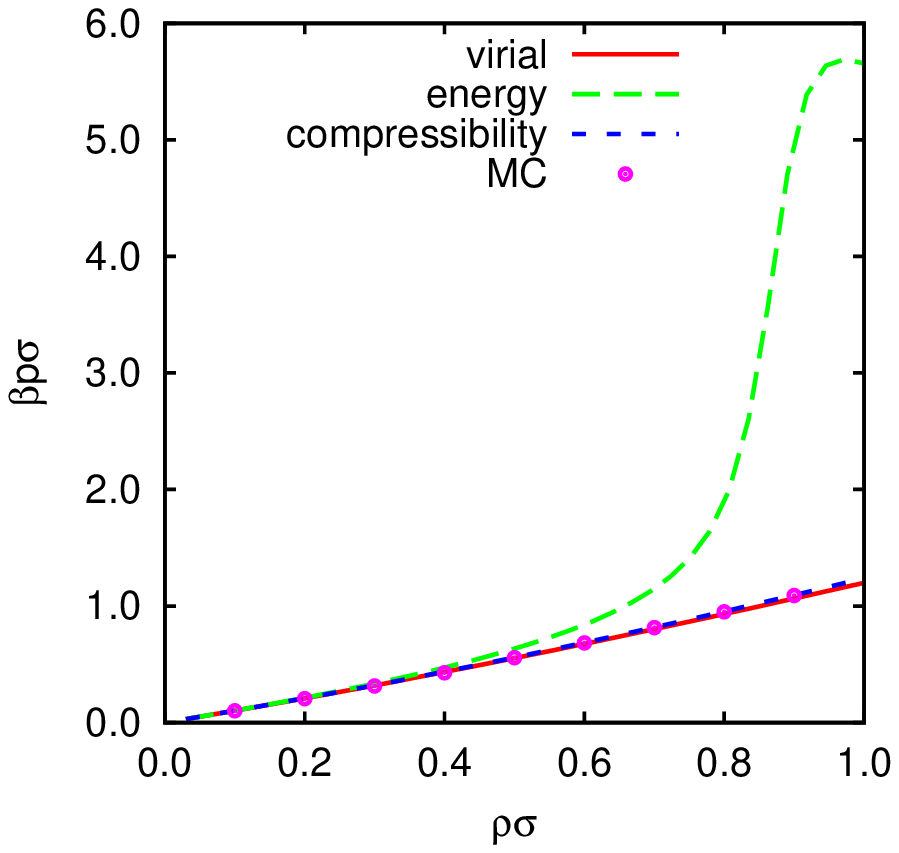}
\end{figure}
\begin{center}
FIG.\ \ref{fig:fig7}. Fantoni \textit{et al.} (JCP)
\end{center}

\clearpage

%
%
\begin{figure}[tbp]
\includegraphics[width=14cm]{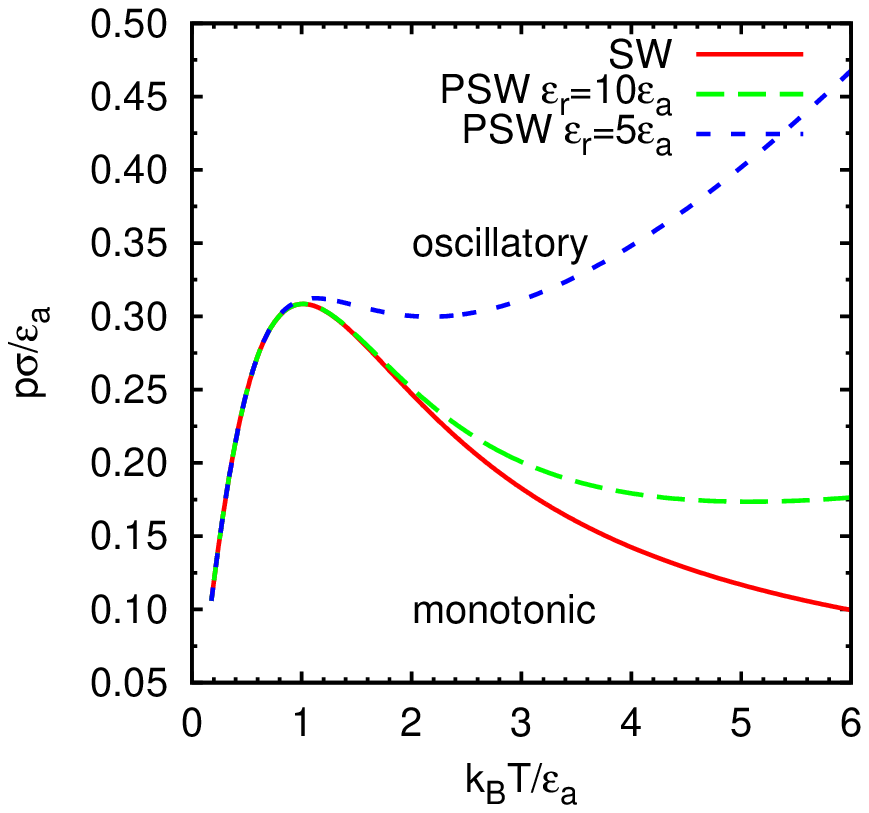}
\includegraphics[width=14cm]{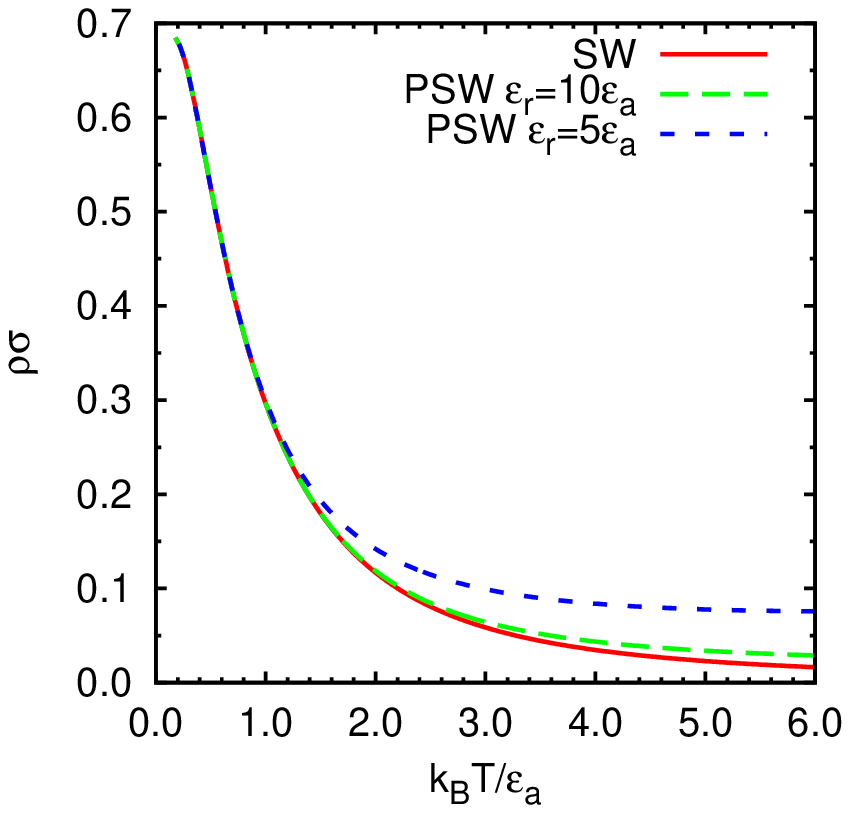}
\end{figure}
\begin{center}
FIG.\ \ref{fig:fig8}. Fantoni \textit{et al.} (JCP)
\end{center}

\clearpage

%
%
\begin{figure}[tbp]
\includegraphics[width=14cm]{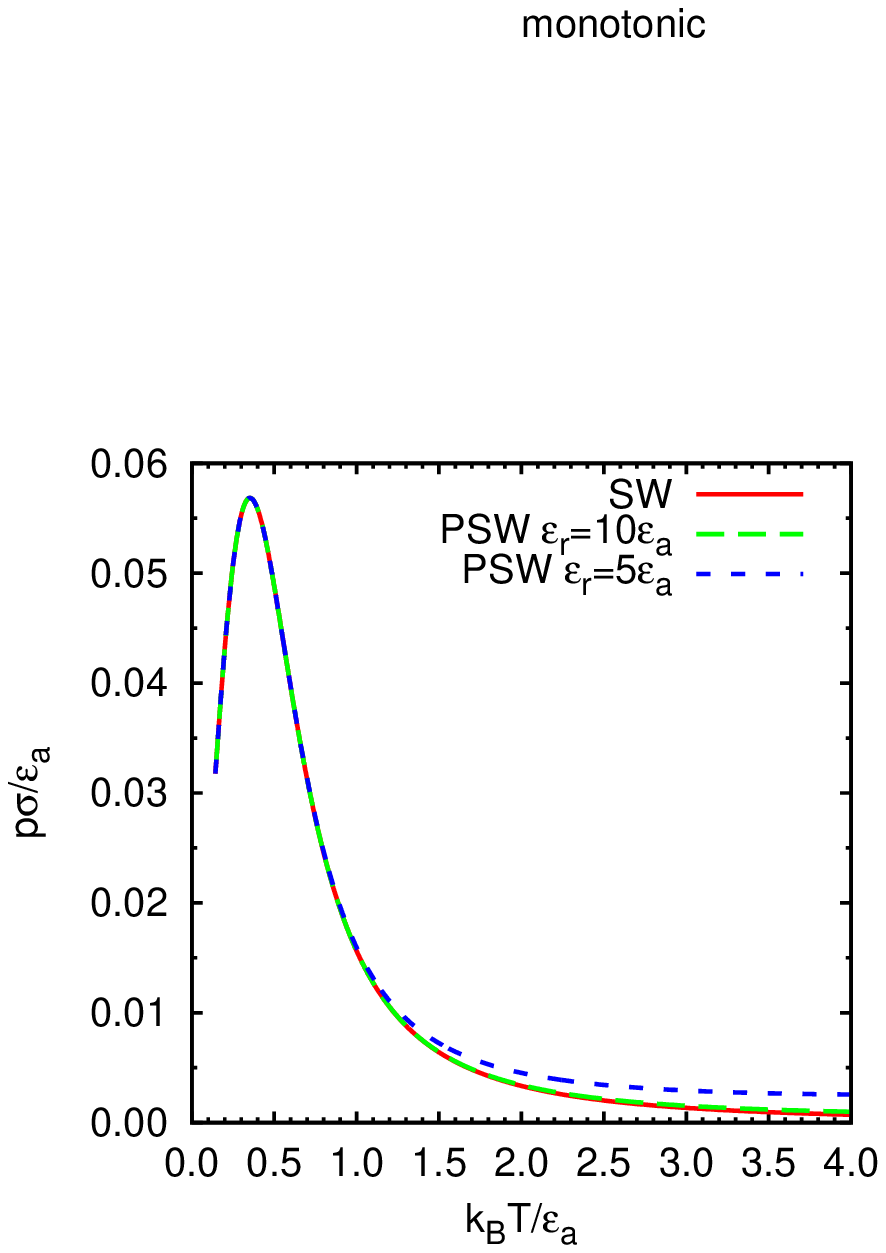}
\includegraphics[width=14cm]{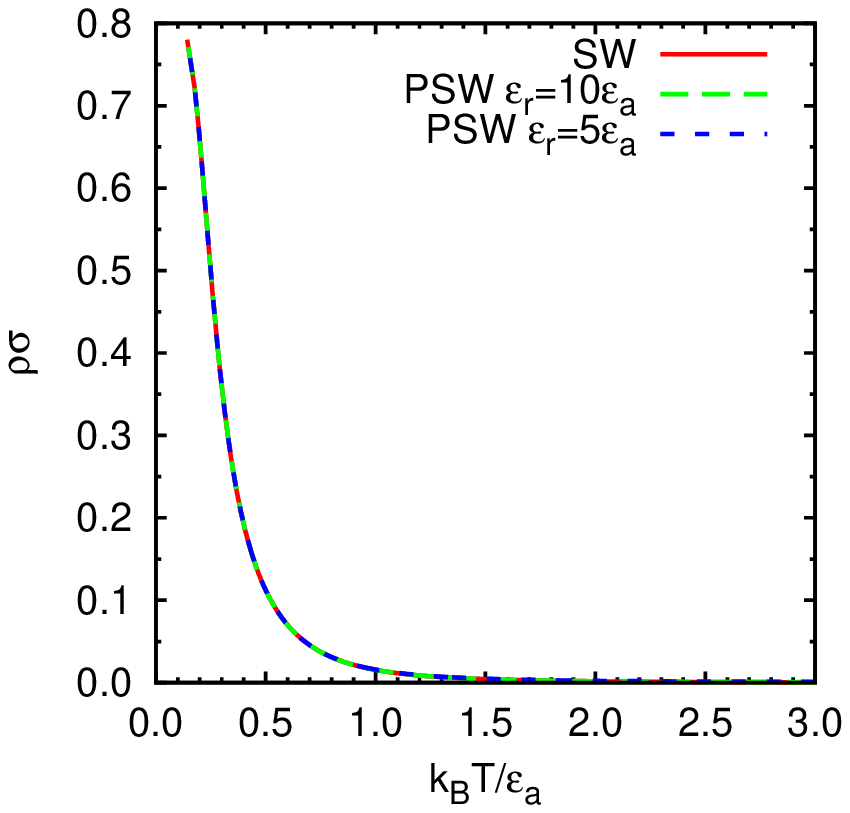}
\end{figure}
\begin{center}
FIG.\ \ref{fig:fig9}. Fantoni \textit{et al.} (JCP)
\end{center}

%

\end{document}